\documentclass[useAMS,usenatbib]{mn2e}
\usepackage{epsfig}

\usepackage{lipsum}
\usepackage{lscape}
\usepackage{longtable,amsmath}
\usepackage{booktabs}
\usepackage{marvosym}
\usepackage{pdflscape}
\usepackage[none]{hyphenat}
\usepackage[T1]{fontenc}
\usepackage{aecompl}
\usepackage{lscape}

\newcommand{\htwo}{H$_2$\,}

\newcommand{\hii}{H{\sc ii}\,}
\newcommand{\fdeg}{$^{\circ}$\!\!}
\newcommand{\fasec}{$\arcsec$\!\!}

\title[The UWISH2 extended \htwo\ source catalogue]{Extended \htwo\ emission line sources from UWISH2}

\author[D.~Froebrich et
al.]{\parbox{\textwidth}{
D.~Froebrich$^{1}$\thanks{E-mail: df@star.kent.ac.uk},
S.V.~Makin$^{1}$,
C.J.~Davis$^{2,3}$, 
T.M.~Gledhill$^{4}$, 
Y.~Kim$^{5}$,
B.-C.~Koo$^{5}$,
J.~Rowles$^{1}$,
J.~Eisl\"offel$^{6}$, 
J.~Nicholas$^{1}$,
J.J.~Lee$^{7}$, 
J.~Williamson$^{1}$ 
A.S.M.~Buckner$^{1}$ 
 \vspace{0.4cm}} \\
$^{1}$Centre for Astrophysics \& Planetary Science, The University of Kent, Canterbury, Kent CT2 7NH, UK \\ 
$^{2}$Division of Astronomical Sciences, National Science Foundation, 4201 Wilson Boulevard, Arlington, VA 22230, USA \\ 
$^{3}$Astrophysics Research Institute, Liverpool John Moores University, Liverpool L3 5RF, UK \\ 
$^{4}$Centre for Astrophysics Research, University of Hertfordshire, College Lane, Hatfield AL10 9AB, UK \\ 
$^{5}$Department of Physics and Astronomy, Seoul National University, Seoul 151-747, Korea \\ 
$^{6}$Th\"uringer Landessternwarte, Sternwarte 5, 07778 Tautenburg, Germany \\
$^{7}$Korea Astronomy and  Space Science Institute, Daejeon 305-348, Korea \\ 
}  

\begin{document}

\date{Accepted. Received.}

\pagerange{\pageref{firstpage}--\pageref{lastpage}} \pubyear{2013}

\maketitle

\label{firstpage}
\begin{abstract}

We present the extended source catalogue for the UKIRT Widefield Infrared
Survey for \htwo\ (UWISH2). The survey is unbiased along the inner Galactic
Plane  from $l \approx 357\degr$ to $l \approx 65\degr$ and $|b| \le
$\,1\fdeg.5 and covers 209 square degrees. A further 42.0 and 35.5 square
degrees of high dust column density regions have been targeted in Cygnus and
Auriga. We have identified 33200 individual extended \htwo\ features. They have
been classified to be associated with about 700 groups of jets and outflows,
284 individual (candidate) Planetary Nebulae,  30 Supernova Remnants and about
1300  Photo-Dissociation Regions. We find a clear decline of star formation
activity (traced by \htwo\ emission from jets and photo-dissociation regions)
with increasing distance from the Galactic Centre. About 60\,\% of the detected
candidate Planetary Nebulae have no known counterpart and 25\,\% of all
Supernova Remnants have detectable \htwo\ emission associated with them.

\end{abstract}


\section{Introduction}
 
The $\nu =$\,1\,--\,0\,(S1) ro-vibrational line of molecular hydrogen at
2.122\,$\mu$m is particularly bright in warm, dense, molecular environments
(T\,$\sim$\,2000\,K, $n_{\rm H}$\,$\ge$\,$10^3$\,cm$^{-3}$). For this reason,
this near-infrared line has been a much-used tracer of shocked molecular gas for
a range of astrophysical phenomena, not least in outflows from the youngest
protostars (e.g. \citet{1995A&A...300..851D}; \citet{2002A&A...392..239S};
\citet{2009A&A...496..153D}; \citet{2010MNRAS.404..661V};
\citet{2012MNRAS.421.3257I}; \citet{2014AJ....148..120B};
\citet{2015AJ....149..101H}; \citet{2015arXiv150608372Z}; Wolf-Chase et al.
(2015, in prep.). \htwo\ may also be excited in photo-dissociation regions
(PDRs) associated with young, intermediate-mass stars and \hii\  regions
(through fluorescence), as well as in post-AGB winds associated with Planetary
and Proto-Planetary Nebulae (PNe; in shocks or again via fluorescence) or in
Supernova Remnants (SNRs).

In late 2006 we defined UWISH2, the UKIRT Wide Field Imaging Survey for \htwo,
as an unbiased, near-infrared, narrow-band imaging survey of the first Galactic
quadrant. The region we initially targeted, covering an area between
10\degr\,$\le$\,$l$\,$\le$\,65\degr\ and
$-$1\fdeg.5\,$\le$\,$b$\,$\le$\,$+$1\fdeg.5, includes most of the giant
molecular clouds and massive star forming regions in the northern hemisphere.
Our goal with UWISH2 was to complement existing and proposed near-, mid- and
far-infrared photometric surveys such as the Spitzer Space Telescope GLIMPSE
survey (\citet{2003PASP..115..953B}; \citet{2009PASP..121..213C}), the Galactic
Plane Survey (GPS, \citet{Lucas2008}) of the UKIRT Infrared Deep Sky Survey
(UKIDSS, \citet{2007MNRAS.379.1599L}), the James Clerk Maxwell Telescope
Galactic Plane Survey (Moore et al. 2015, subm.), the Herschel Space Telescope
Hi-Gal survey (\citet{2010PASP..122..314M}), by utilising the \htwo\
1\,--\,0\,S(1) line as a tracer of the {\em dynamically active} component of
star formation (SF) not  emphasised by the broad-band surveys.

Much of the UWISH2 survey area has also recently been imaged with the same
telescope and instrument in narrow-band [FeII] line emission at 1.64\,$\mu$m
(the UKIRT Wide Field Infrared Survey for Fe$^{+}$ -- UWIFE;
\citet{2014MNRAS.443.2650L}).  These observations are certainly complementary to
the \htwo\ imaging presented here since [FeII] is an excellent tracer of the
higher-excitation atomic gas in shocks and collimated jets (e.g.
\citet{2002A&A...393.1035N}; \citet{2002ApJ...570L..33G};
\citet{2004A&A...419..999G}).

The UWISH2 survey was completed in 2011 and is described in
\citep{2011MNRAS.413..480F}.  An extension to the survey, referred to as
UWISH2-E, was proposed in late 2012.  Between December 2012 and December 2013,
our large mosaic of \htwo\ images of the Galactic Plane (GP) was extended down
through the Galactic Center to $l$\,$\sim$\,357\degr\ (although this extension
does not cover the full width of the original survey at all longitudes - see
Sect.\,\ref{areas}).  We also partially mapped two new fields, one available in
the summer, the other in winter, around the well-known high mass star forming
regions in Cygnus and the more quiescent molecular cloud complex in Auriga.

In this paper we present the results of an unbiased search for all extended
\htwo\ emission line features in the UWISH2 and UWISH2-E surveys. We aim to
provide a comprehensive catalogue of extended 1\,--\,0\,S(1) features, their
properties (position, size and flux) and most likely classification (as
jet/outflow, PN, SNR or unclassified). This catalogue will be useful as a
starting point for more detailed investigations of selected sub-sets of \htwo\
emission line objects, such as individual jets and outflows or PNe. 

In Sect.\,\ref{survey} we describe the observations, survey areas, and data
quality. In Sect.\,\ref{catalogue} we present the extended source catalogue and
give a detailed account of the techniques used to find emission line features in
the survey images. In Sect.\,\ref{discussion} we discuss the overall properties
of the detected \htwo\ emission line features, but refer to future publications
for the detailed study of selected individual objects.


\section{The UWISH2 and UWISH2-E Surveys}\label{survey}

\subsection{Observations}

All data were acquired using the Wide Field Camera (WFCAM) on the United
Kingdom Infrared Telescope (UKIRT), Mauna Kea, Hawaii. WFCAM houses four
Rockwell Hawaii-II (HgCdTe 2048\,$\times$\,2048\,pixel) arrays spaced by
94\,\% in the focal plane. The pixel scale measures 0\fasec.4, although
micro-stepping is used to generate reduced mosaics with a 0\fasec.2 pixel
scale and thereby fully sample the expected seeing.

For both the UWISH2 and UWISH2-E surveys we essentially repeated the observing
strategy adopted by the UKIDSS GPS \citep{Lucas2008}, the only difference being
the choice of filter and the exposure time used.  Individual 60\,s exposures
through a narrowband \htwo\ filter ($\lambda = 2.122$\,$\mu$m, $\delta\lambda =
0.021$\,$\mu$m) were repeated with a 2\,$\times$\,2 point micro-stepping at
three jitter positions. In this way 12 exposures were acquired at each telescope
pointing, resulting in a total exposure time per pixel of 720\,s.  Four
telescope pointings are needed to fill in the gaps between the detectors; 16
mosaic images thus constitute a tile covering about 0.75 square degrees.

All data were reduced by the Cambridge Astronomical Survey Unit (CASU), which is
responsible for data processing prior to archiving and distribution by the Wide
Field Astronomy Unit (WFAU). The CASU reduction steps are described in detail by
\citet{2006MNRAS.372.1227D}; astrometric and photometric calibrations were
achieved using 2MASS (\citet{2006MNRAS.372.1227D}; \citet{2006MNRAS.367..454H}).
The reduced images are available from WFAU as well as from the UWISH2
website\footnote{Data available from http://astro.kent.ac.uk/uwish2/}, along
with the corresponding broad-band J, H and K images from the GPS data.
Continuum-subtracted \htwo$-$K images are also available, as are colour
renditions of each 16-image tile.

\begin{figure}
\includegraphics[width=8.6cm,angle=0]{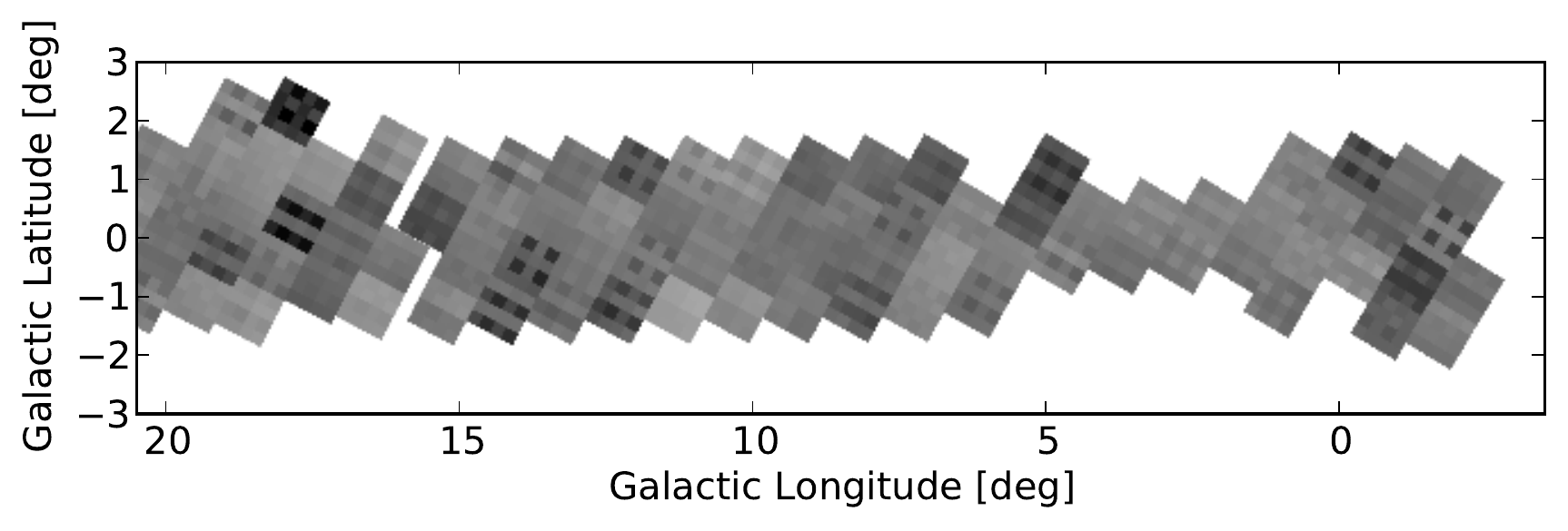} \\
\includegraphics[width=8.6cm,angle=0]{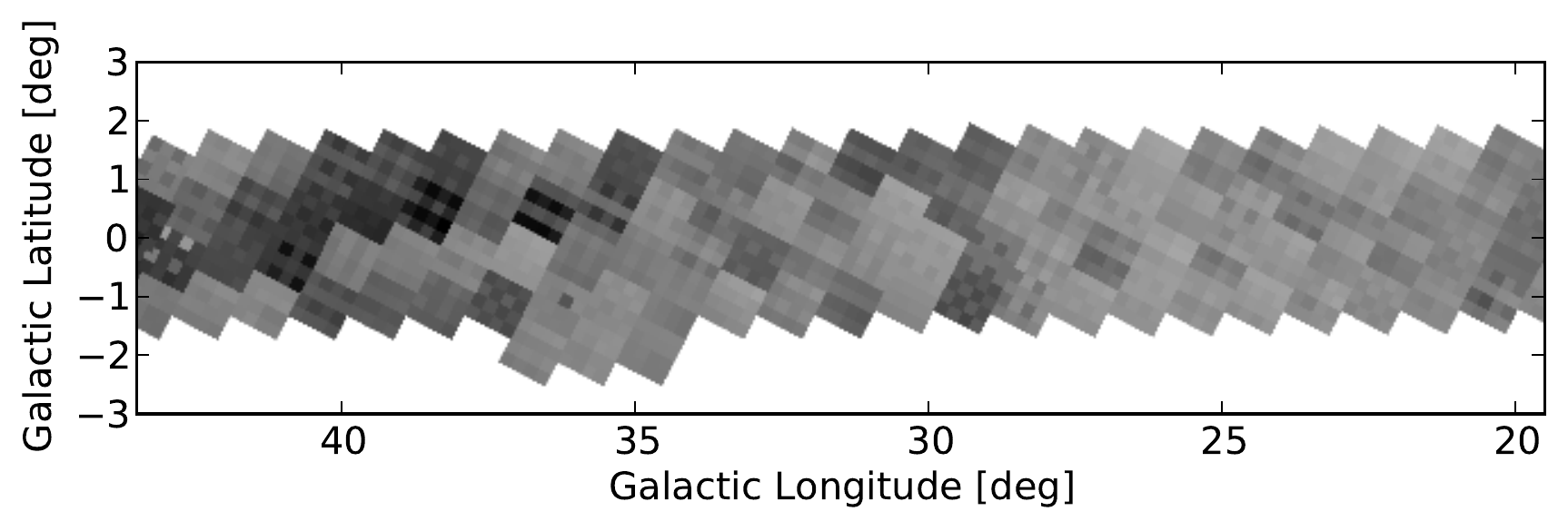} \\
\includegraphics[width=8.6cm,angle=0]{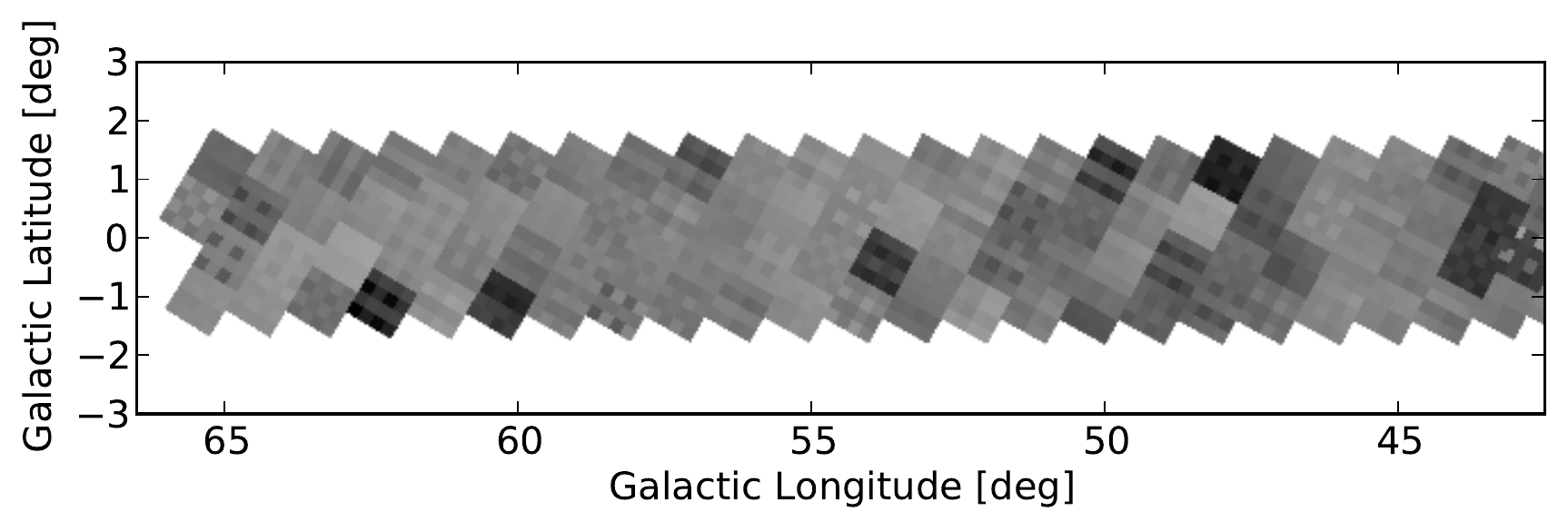} \\

\caption{\label{gpplots}  Plots of the seeing distribution in the Galactic Plane
area of the survey. Positions covered by tiles/images with worse seeing are
indicated by darker colours. See Sect.\,\ref{characteristics} for more details.}

\end{figure}

\begin{figure}
\includegraphics[width=8.6cm,angle=0]{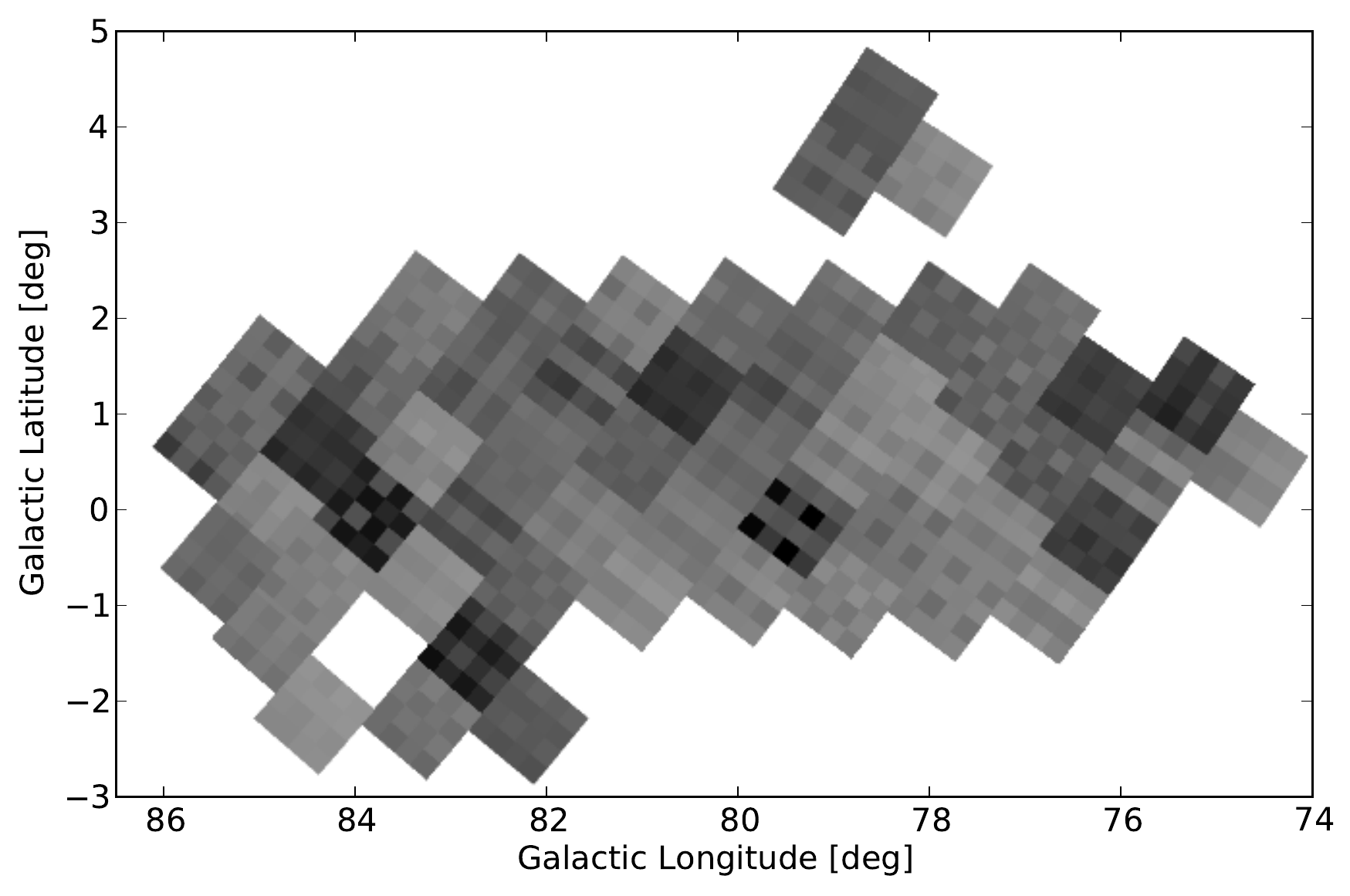} \\
\includegraphics[width=8.6cm,angle=0]{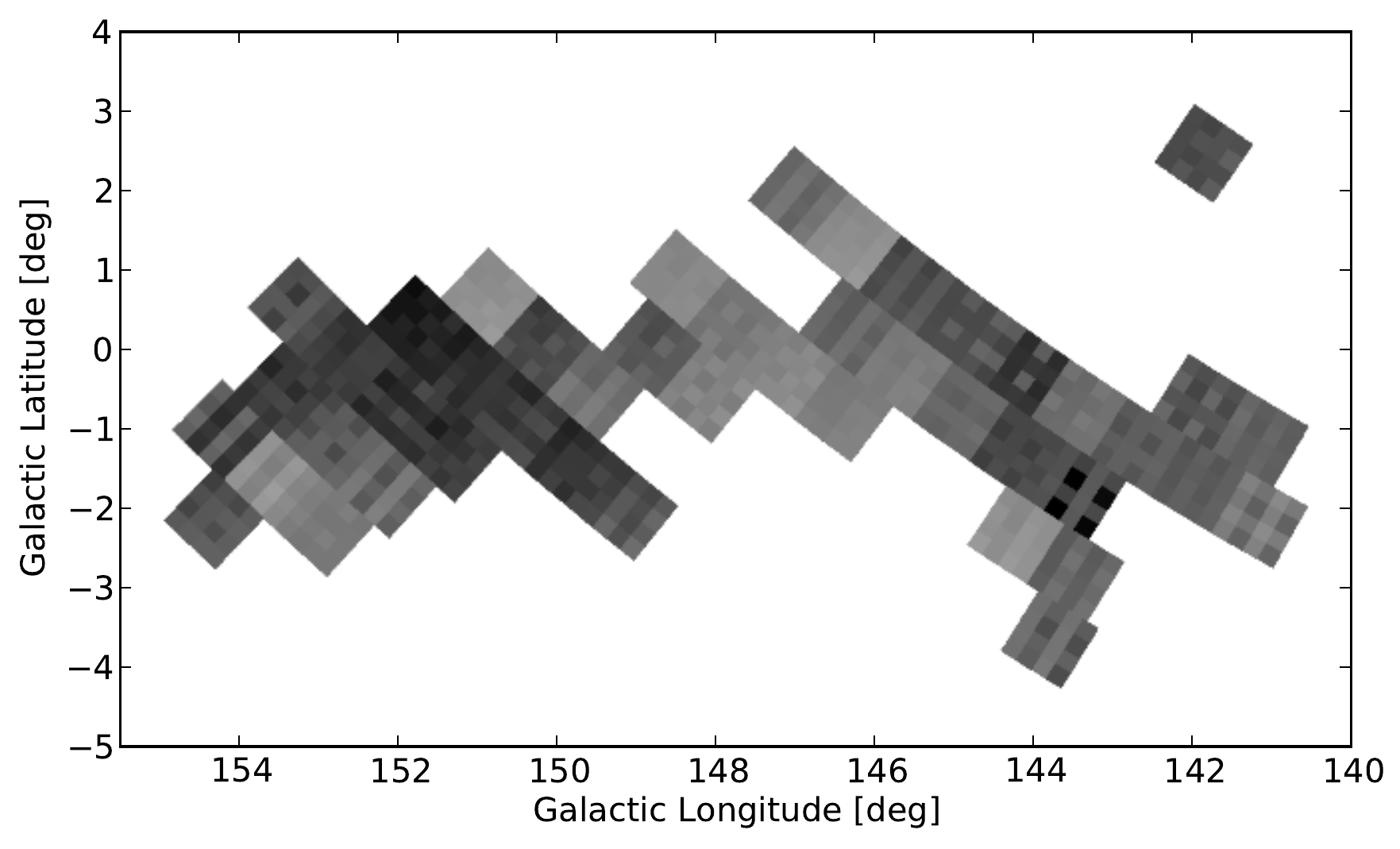} \\

\caption{\label{cygaurplots}  As Fig.\,\ref{gpplots} but for the Cygnus (top)
and Auriga (bottom) area of the survey. }

\end{figure}

\subsection{Target Area}\label{areas}

The survey covers the northern GP as well as selected high dust column density
regions in Cygnus and Auriga. Along the GP we covered a longitude range from $l
\approx 357\degr$ to about $l \approx 65\degr$. For most of this longitude range
the survey covers the region $|b|$\,$\le $\,1\fdeg.5. There are some extensions
towards the North at $l$\,$\approx$\,$19\degr$ and towards the South at
$l$\,$\approx$\,$36\degr$. Furthermore, due to time constraints we were unable
to complete the full latitude range near the Galactic Centre.
Figures\,\ref{gpplots} and \ref{cygaurplots} show the detailed coverage along
the GP and in Cygnus and Auriga. In total we have observed 268 tiles with
coverage in \htwo\ and the UKIDSS GPS K-band. Note that we have observed one
additional tile just South of the Galactic Centre, but there is no K-band
counterpart in the GPS database. We have searched the image for \htwo\ emission,
but no features were detected. 

Considering the overlap between images and tiles, the total area covered along
the GP is 209 square degrees.  In the Cygnus and Auriga areas the fields are
roughly concentrated along the GP, but preference has been given to high
extinction regions. Full coverage of the entire cloud complex could not be
obtained due to time constraints. We have observed 54 tiles in Cygnus and 45
tiles in Auriga. Considering the overlap of images, this corresponds to 42.0 and
35.5 square degrees, respectively. Hence the total area covered in the entire
UWISH2 survey is 286.5 square degrees.

\begin{figure*}
\includegraphics[width=5.6cm,angle=0]{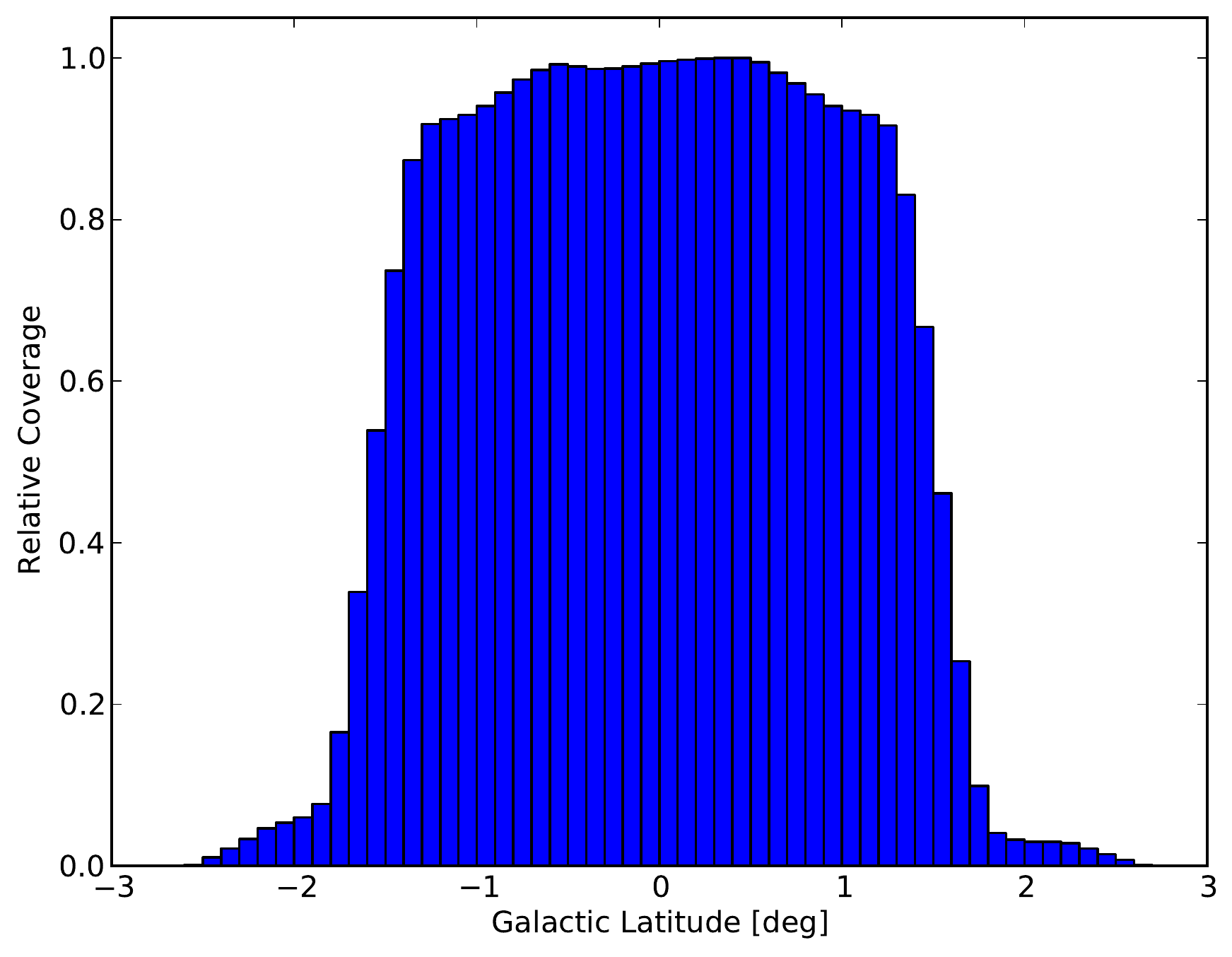} \hfill
\includegraphics[width=5.6cm,angle=0]{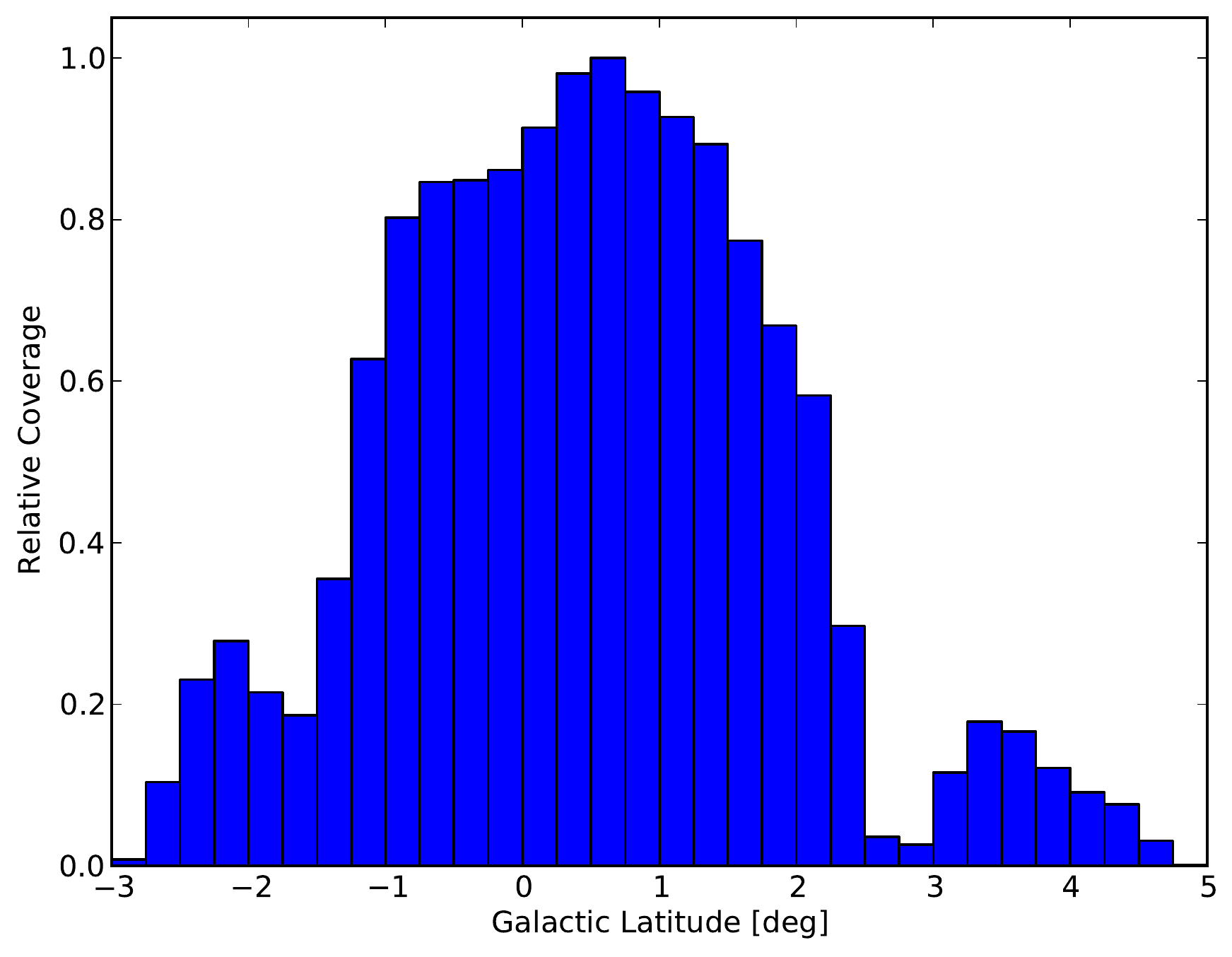} \hfill
\includegraphics[width=5.6cm,angle=0]{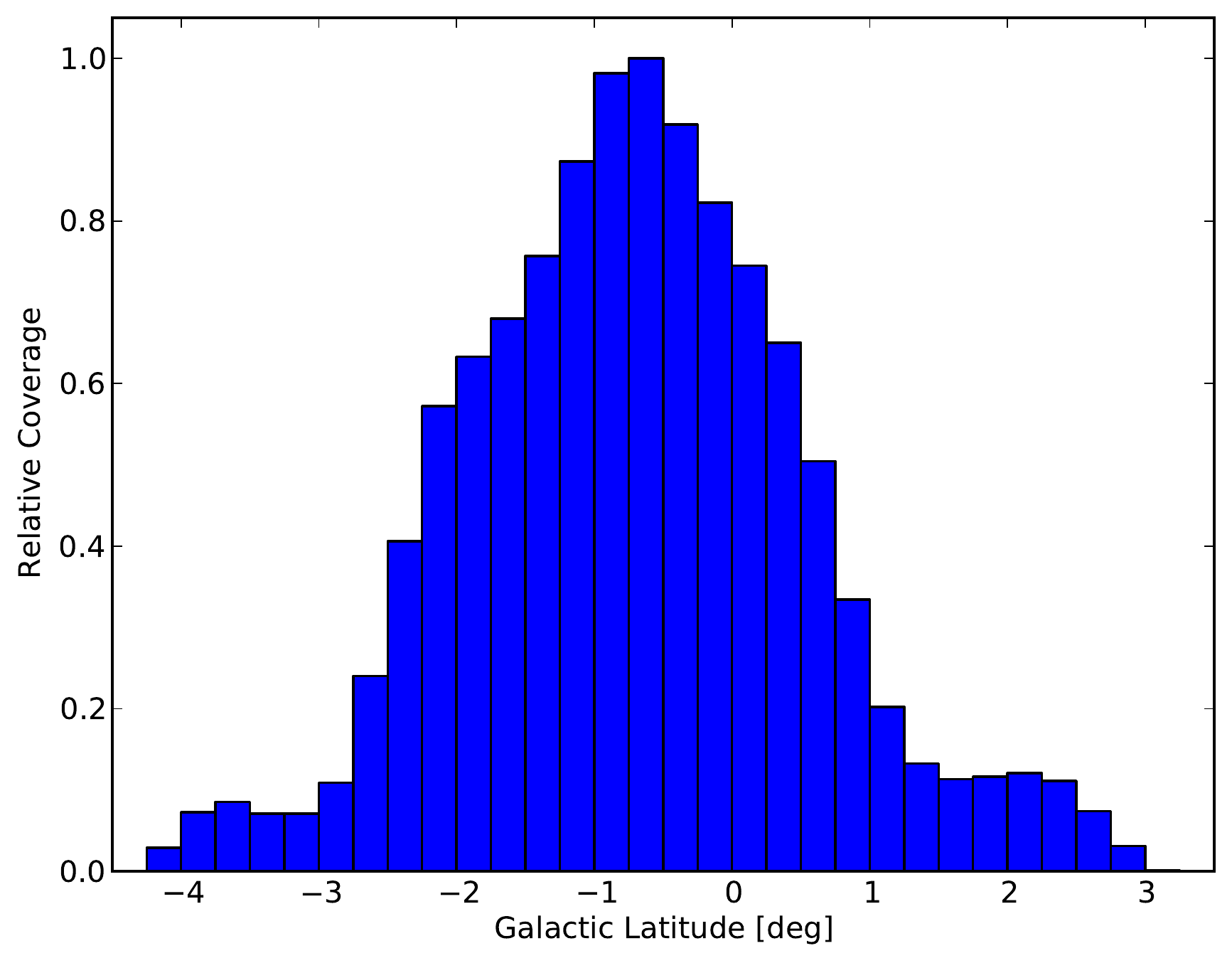} \\
\includegraphics[width=5.6cm,angle=0]{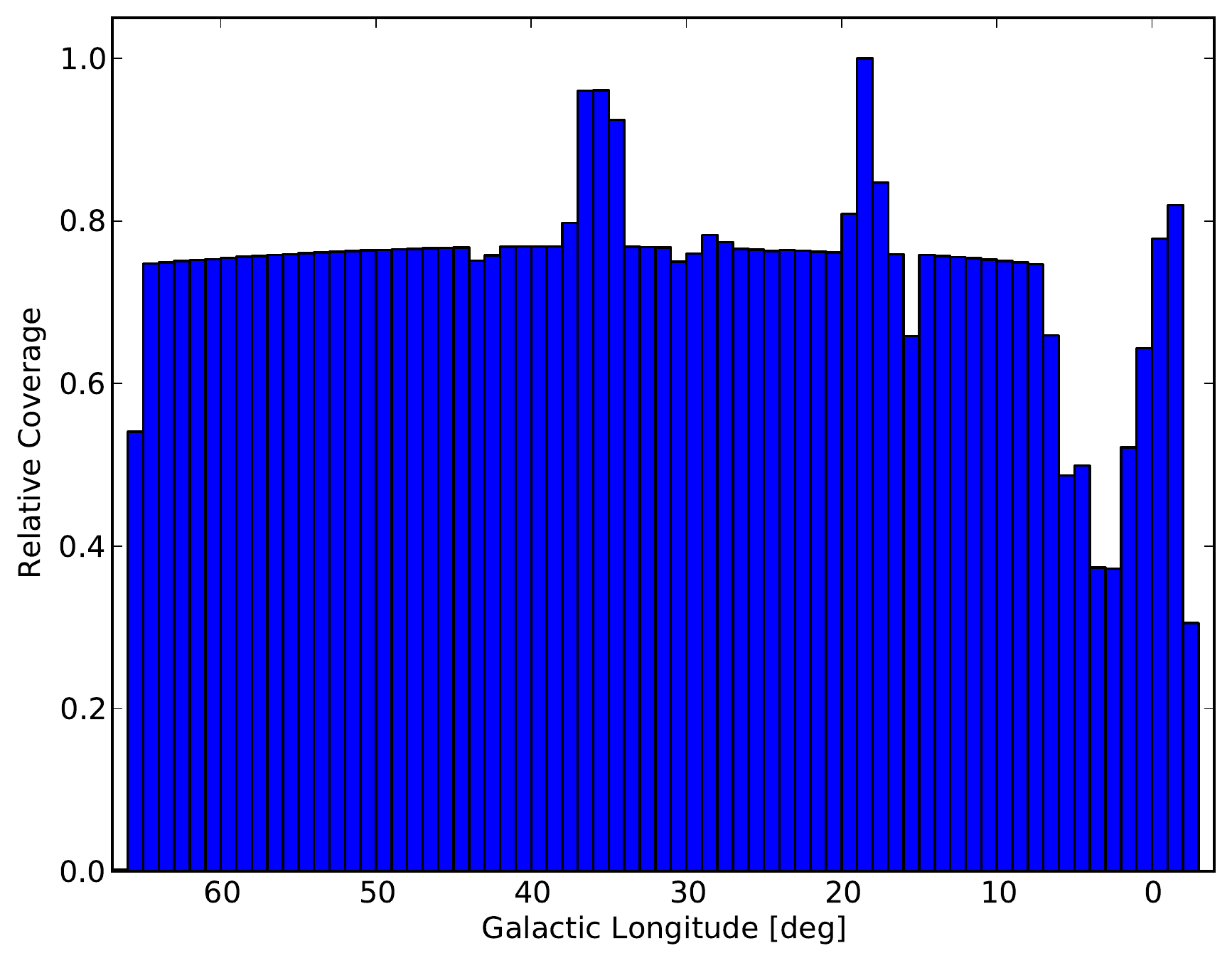} \hfill
\includegraphics[width=5.6cm,angle=0]{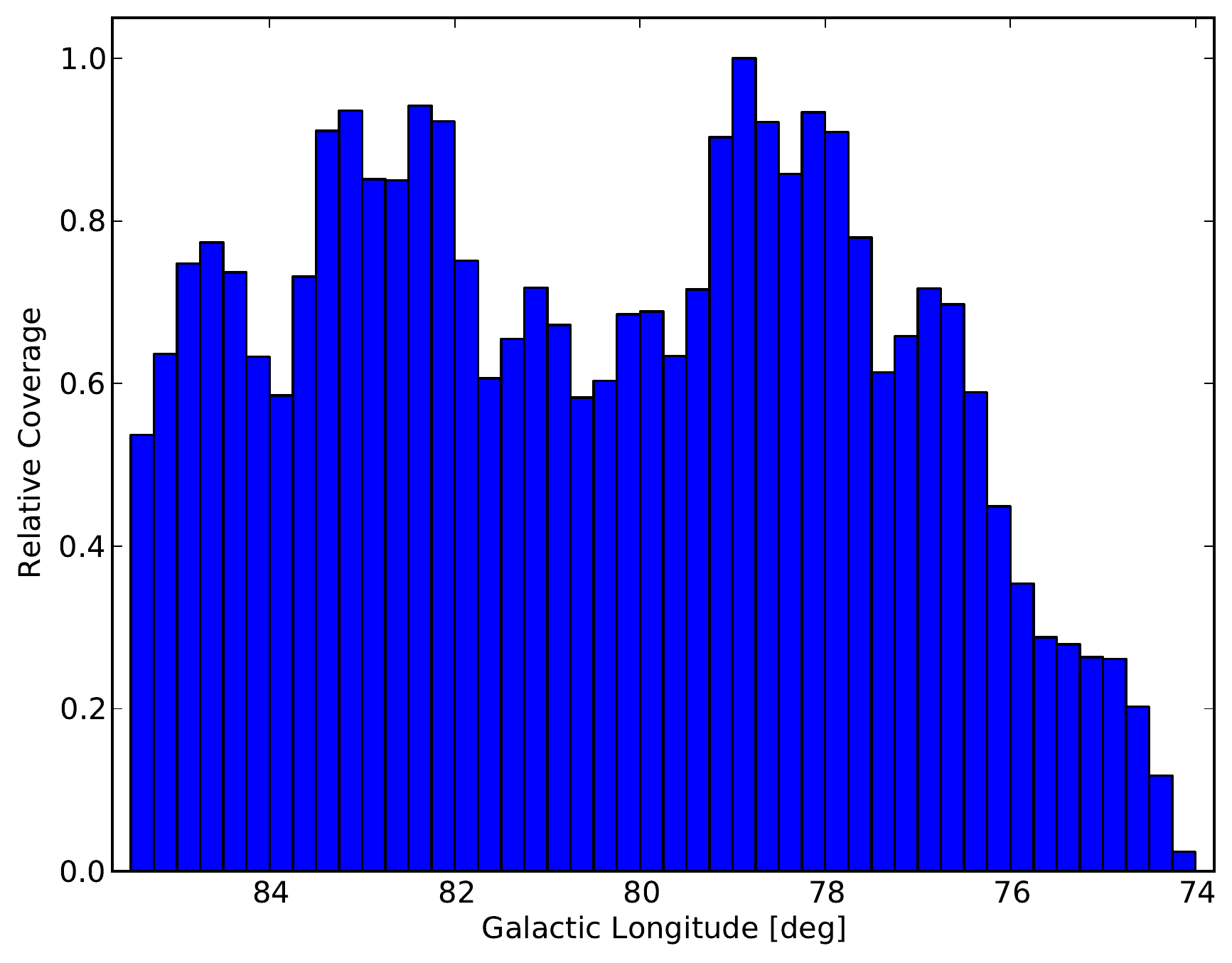} \hfill
\includegraphics[width=5.6cm,angle=0]{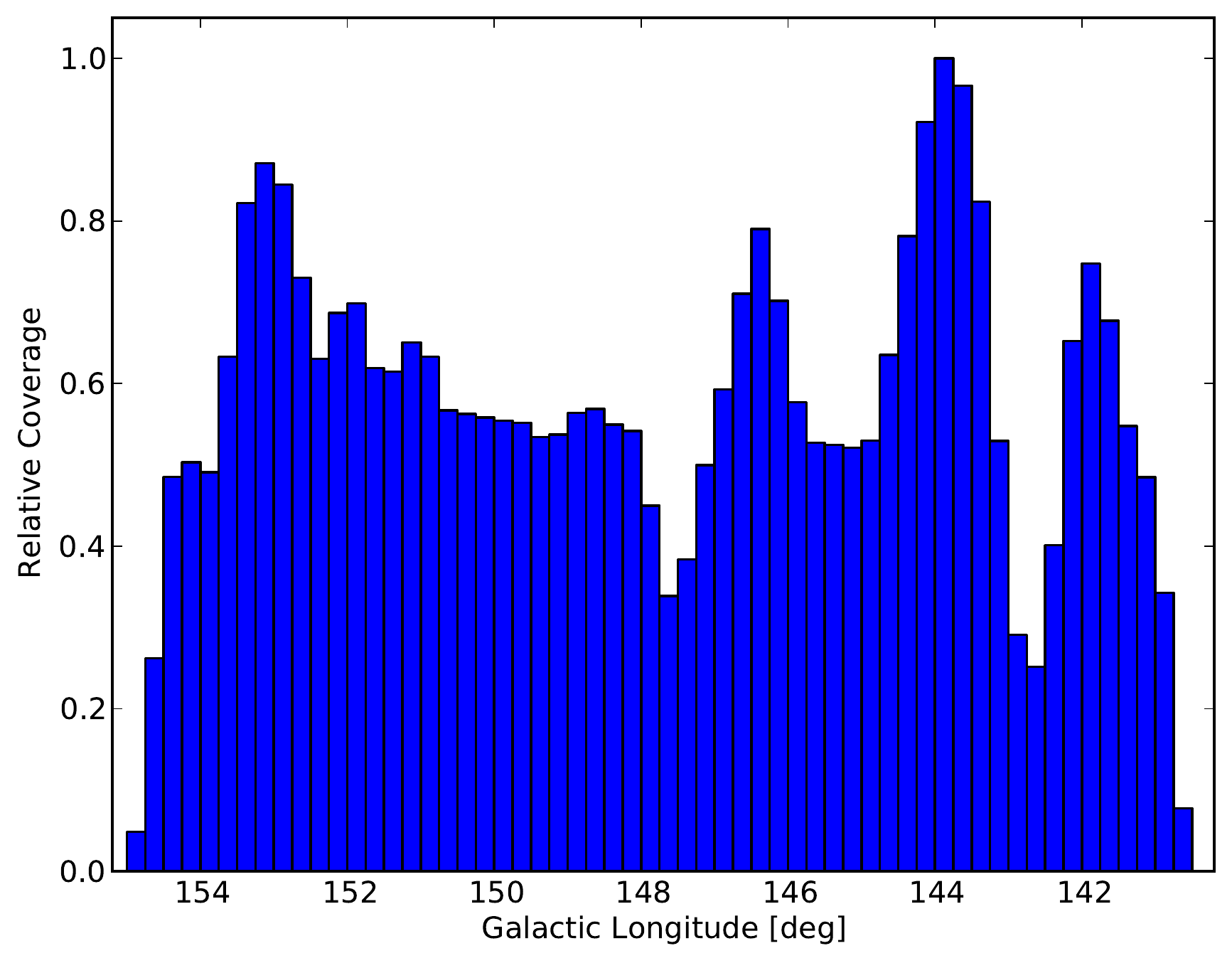} \\

\caption{\label{covdist}  Relative latitude (top row) and longitude (bottom row)
coverage of the survey in the Galactic Plane (left column), Cygnus (middle
column) and Auriga (right column) area of the survey. The relative coverage is
proportional to the number of images taken at a specific latitude/longitude and
is normalised to a maximum of one. These distributions are used to correct
observed distributions of objects such as Jets and PNe to account for the
variations in coverage along and across the GP. }

\end{figure*}

Due to the nature of the observations, the coverage in both Galactic longitude
and latitude in the survey area is not homogeneous. Hence any investigations of
distributions of objects along and perpendicular to the GP will have to be
corrected for the variations in relative coverage, i.e. a factor proportional to
the number of images obtained at a specific latitude/longitude. In the top left
panel of Fig.\,\ref{covdist} we show the relative coverage (normalised to a
maximum of one) perpendicular to the GP. As can be seen, within 1\fdeg.3 of the
GP, the relative coverage exceeds 90\,\% and is more or less homogeneous.
Further away from the GP the coverage steeply declines and sinks below 10\,\% at
about  1\fdeg.8 from the GP. A  50\,\% coverage is achieved for all areas with
$|b| < $\,1\fdeg.5. In the bottom left panel of Fig.\,\ref{covdist} we show a
similar graph for the coverage along the GP. Between $l = 7\degr$ and $l =
65\degr$ the coverage is almost constant. Larger discrepancies are only seen
near the GC and the two areas where we observed additional tiles slightly
further away from the GP (at $ l \approx 18\degr$ and $l \approx 35\degr$). For
completeness, we also show the coverage distributions for the Cygnus and Auriga
regions in the middle and right columns of Fig.\,\ref{covdist}, respectively.
Due to the more patchy distribution of tiles in these clouds, the relative
coverage in these cases is much more variable than along the GP.

 %
 %

\begin{figure*}
\includegraphics[width=5.6cm,angle=0]{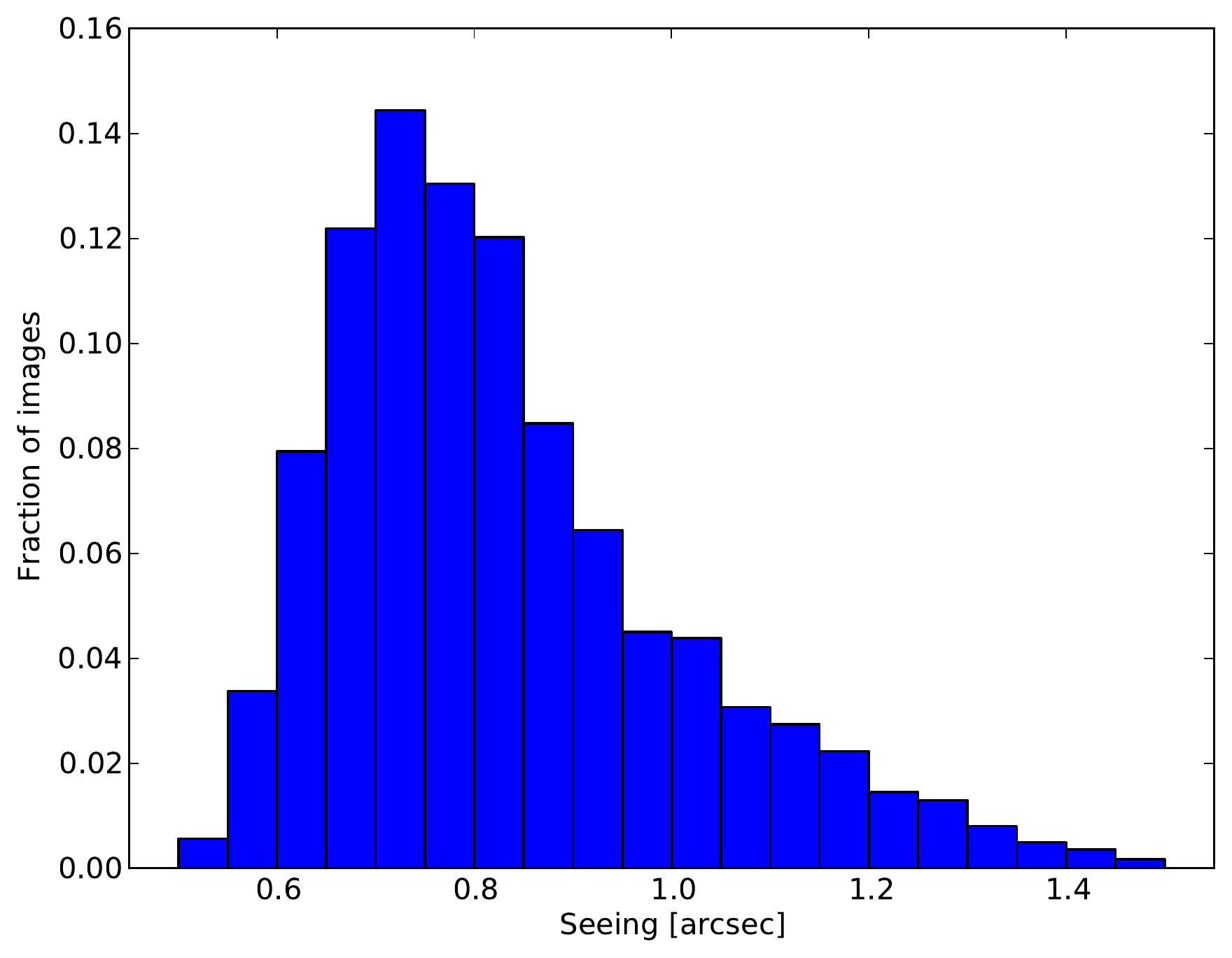} \hfill
\includegraphics[width=5.6cm,angle=0]{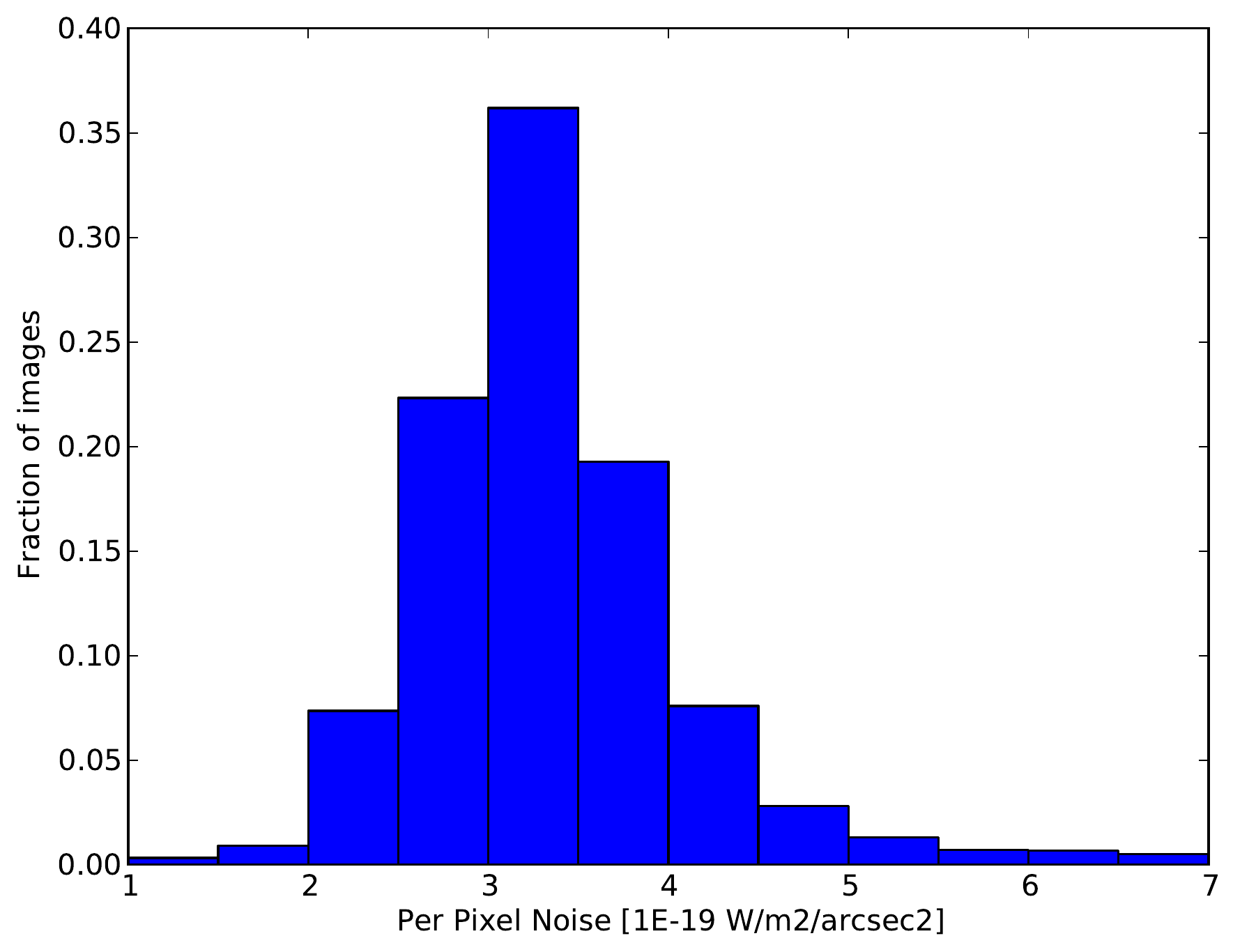} \hfill
\includegraphics[width=5.6cm,angle=0]{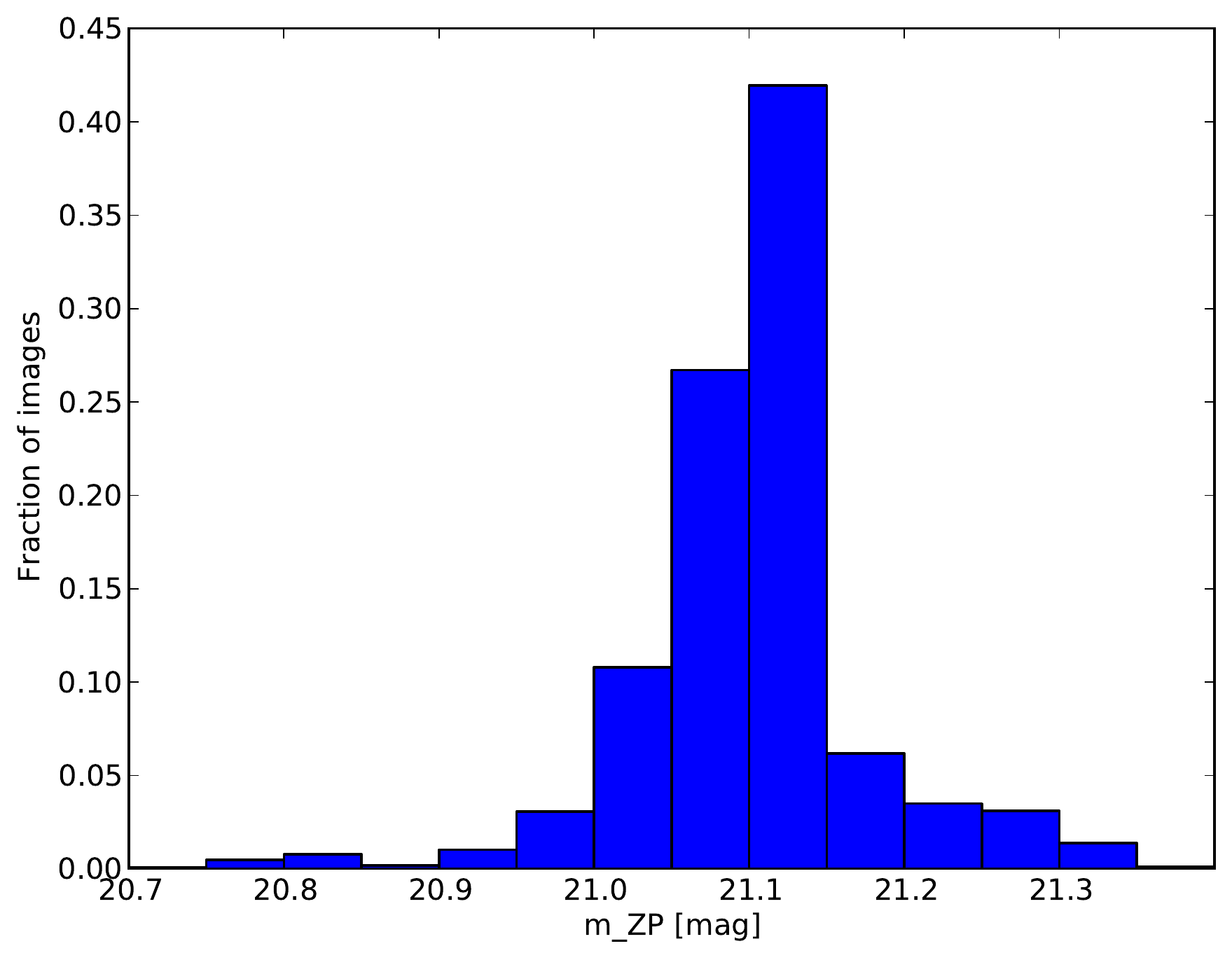} \\

\caption{\label{seeinghist}  Data quality of the survey. The left panel shows
the seeing distribution of our images , the middle panel the one sigma per pixel
noise distribution and the right panel the distribution of the photometric zero
point {\tt mag$\_$zp}. The median seeing is 0\fasec.8 and about 83\,\% of the
data has been taken with a seeing of less than one arcsecond. The median one
sigma per pixel noise is
3.25\,$\times$\,$10^{-19}$\,W\,m$^{-2}$\,arcsec$^{-2}$.}

\end{figure*}

 %
 %
 %
 %

\subsection{Data Characteristics}\label{characteristics}

The distribution of seeing values in our survey can be found in the left panel
of Fig.\,\ref{seeinghist}. How these values are distributed spatially can be
seen in Figs.\,\ref{gpplots} and \ref{cygaurplots}. The median seeing in the
survey is 0\fasec.79, with 82.9\,\% of the area observed at seeing values of
below one arcsecond. Most of the poorer seeing data is distributed in the
additional regions in Cygnus and Auriga. There, however, the crowding of stars
is much less severe than in the inner GP, hence slightly worse seeing will not
affect the detection and photometry of extended \htwo\ features.

We determine the background per pixel noise level in the images by estimating
the {\it rms} scatter of the pixel values from the background, using a 3 sigma
clipping procedure to remove stars. The counts are then converted into a surface
brightness using the {\tt mag$\_$zp} values and integration times (for details
see the calibration of photometry in Sect.\,\ref{calibration}). The distribution
of the one pixel 1$\sigma$ noise for all images is shown in the middle panel of
Fig.\,\ref{seeinghist}. The median one pixel noise is $3.25 \times
10^{-19}$\,W\,m$^{-2}$\,arcsec$^{-2}$, in agreement with the typical noise in
the original UWISH2 area \citep{2011MNRAS.413..480F}. Averaged over the median
seeing from above, which covers about 16 pixels, the typical 5$\sigma$ noise or
surface brightness detection limit is $4.1 \times
10^{-19}$\,W\,m$^{-2}$\,arcsec$^{-2}$. Alternatively, the 3$\sigma$ noise over
1\fasec.2, the Glimpse pixel size, is $1.6 \times
10^{-19}$\,W\,m$^{-2}$\,arcsec$^{-2}$.

In the right panel of Fig.\,\ref{seeinghist} we show the distribution of the
photometric zero point values in the images. The narrow peak around 21.1\,mag
indicates that about two thirds of all images were taken under comparable
atmospheric conditions, with extinction variations of less than 5\,\%. This can
also be seen in the Figs.\,\ref{gpplotsapp} and \ref{cygaurplotsapp} in the
Appendix. We also summarise all the data for every image in the Appendix in
Table\,\ref{imagedatatable}. There we list the tile containing the image, the
image name (containing the observation date), the centre of each image in RA,
DEC (J2000) and l,b, the seeing, the calibration magnitude zero point and its
uncertainty as well as the estimated one pixel surface brightness noise.


\section{The Extended Source Catalogue}\label{catalogue}

In this section we describe the extended \htwo\ emission line object 
catalogue obtained from the UWISH2 images. 

\subsection{Source detection}

To obtain an, as much as possible, complete and unbiased catalogue of extended
\htwo\ emission line features, we performed the following steps for all images:
i) Continuum subtraction of the emission line images; ii) Filtering and
automated detection of extended \htwo\ features; iii) Manual verification and
removal of image artifacts. These steps are performed as described in detail
below.

\subsubsection{Continuum Subtraction}

To remove the continuum emission from the \htwo\ narrow band images we utilised
the K-band data from the UKIDSS GPS \citep{Lucas2008}. This continuum
subtraction was done on an image by image basis, i.e. run separately for each
4k$\times$4k image. Most of our \htwo\ images were taken at exactly the same
positions as the GPS K-band data, with off-sets of less than a fraction of an
arcminute. For a small fraction of fields, the off-sets were larger than one
arcminute. In these cases we combined the K-band images from the GPS to obtain
a matching K-band image utilising the
Montage\footnote{http://montage.ipac.caltech.edu/index.html} software. The
image subtraction routine aligns the \htwo\ and K-band images, determines the
scaling factor for the continuum image and uses psf-fitting to subtract the
stars. The K-band scale factor and the psf shape are determined from
unsaturated, isolated stars in 1000\,pix\,$\times$\,1000\,pix sub-images. The
details of the procedure are described in \citet{2014MNRAS.443.2650L}. Note
that the fluxes in the \htwo\ images are unchanged, and only the K-band
continuum data are scaled. 

\begin{figure}
\includegraphics[width=2.75cm,angle=0]{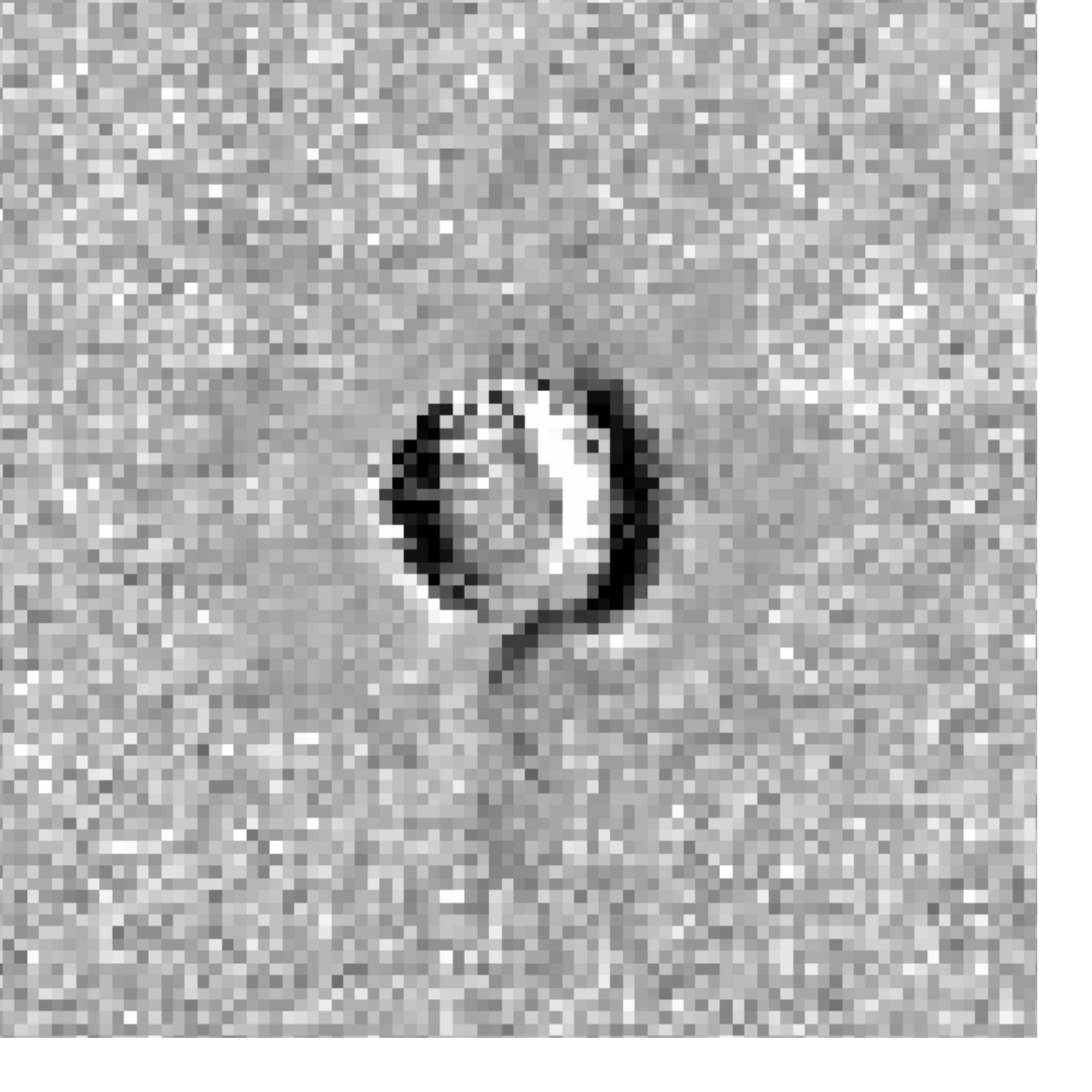} \hfill
\includegraphics[width=2.75cm,angle=0]{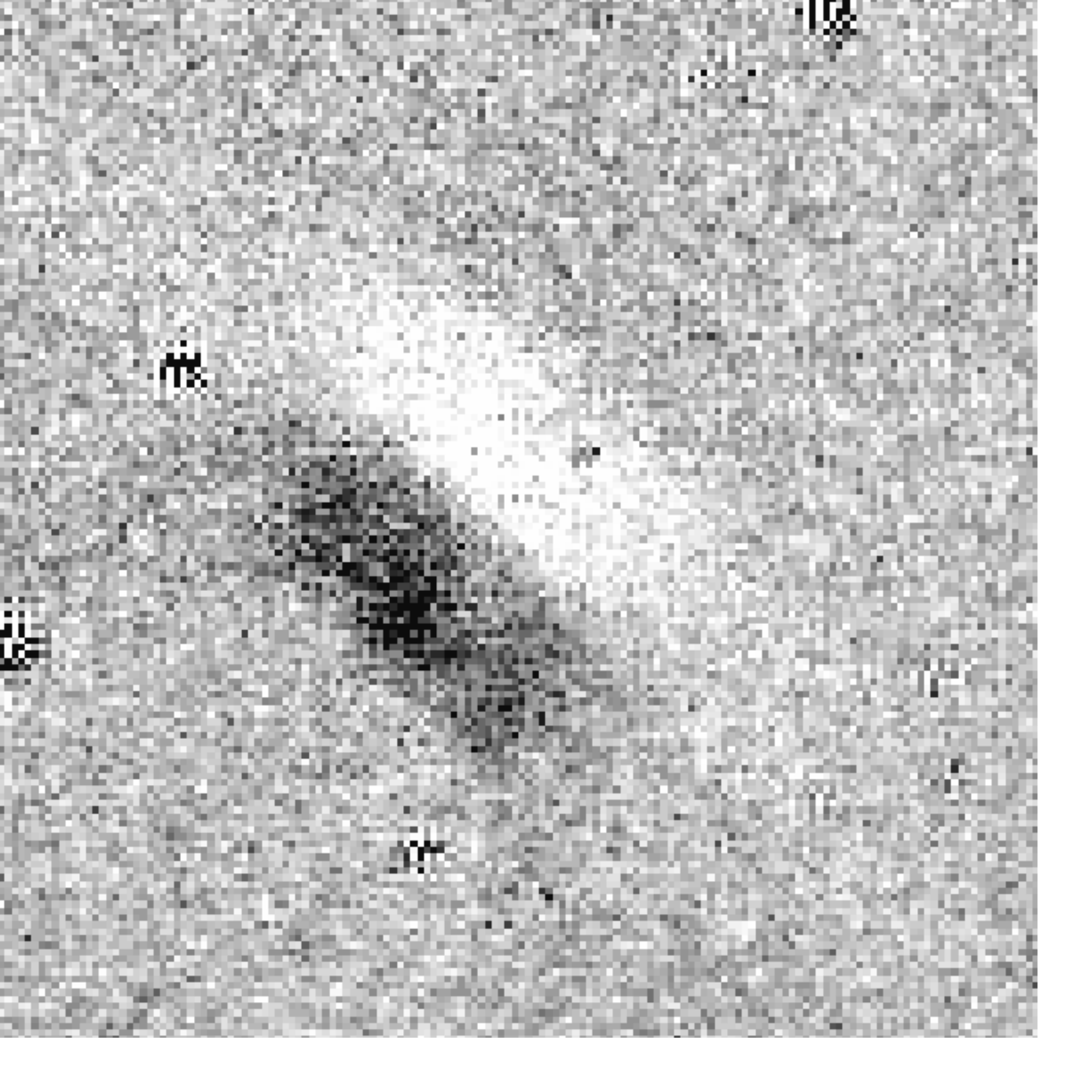} \hfill
\includegraphics[width=2.75cm,angle=0]{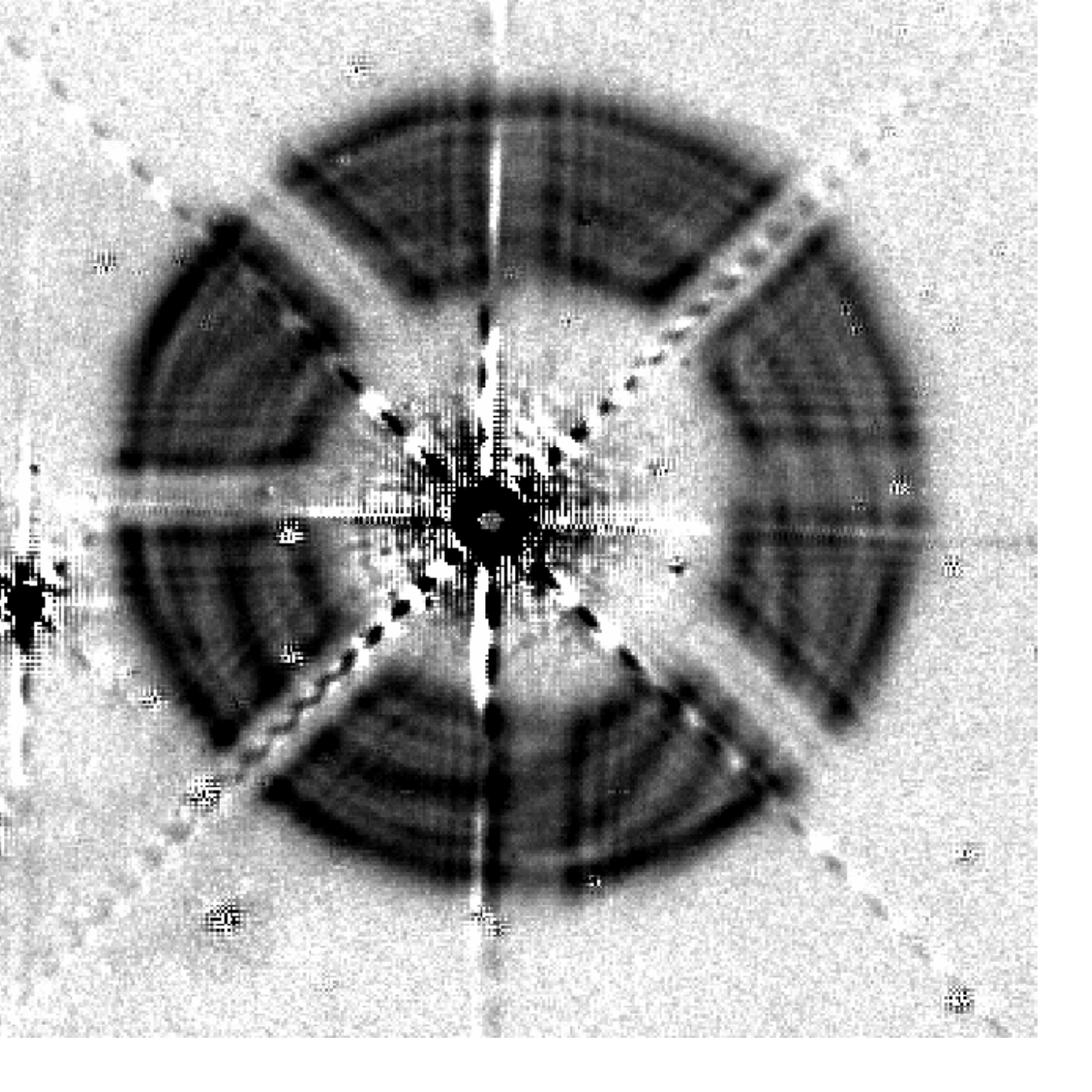} \\
\includegraphics[width=2.75cm,angle=0]{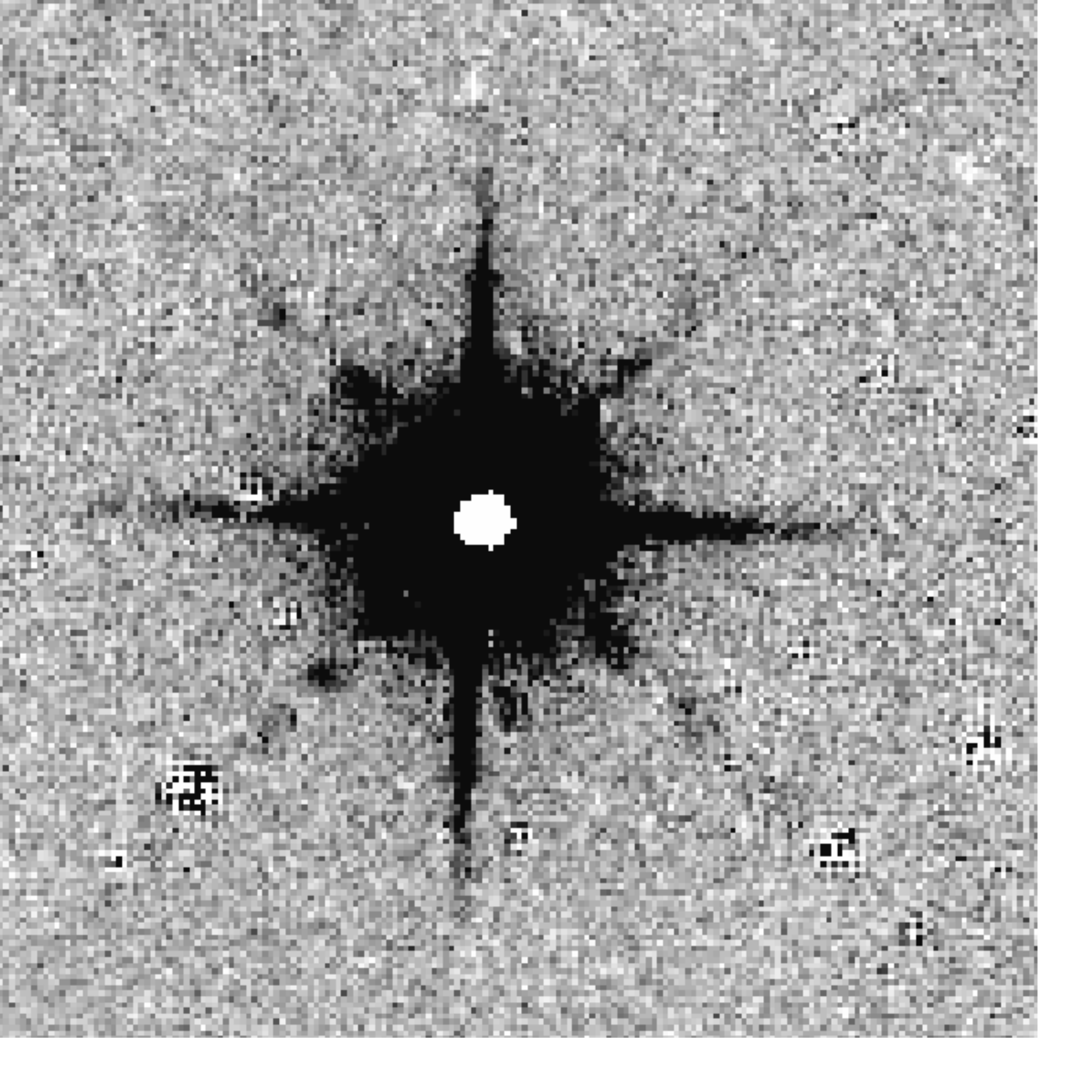} \hfill
\includegraphics[width=2.75cm,angle=0]{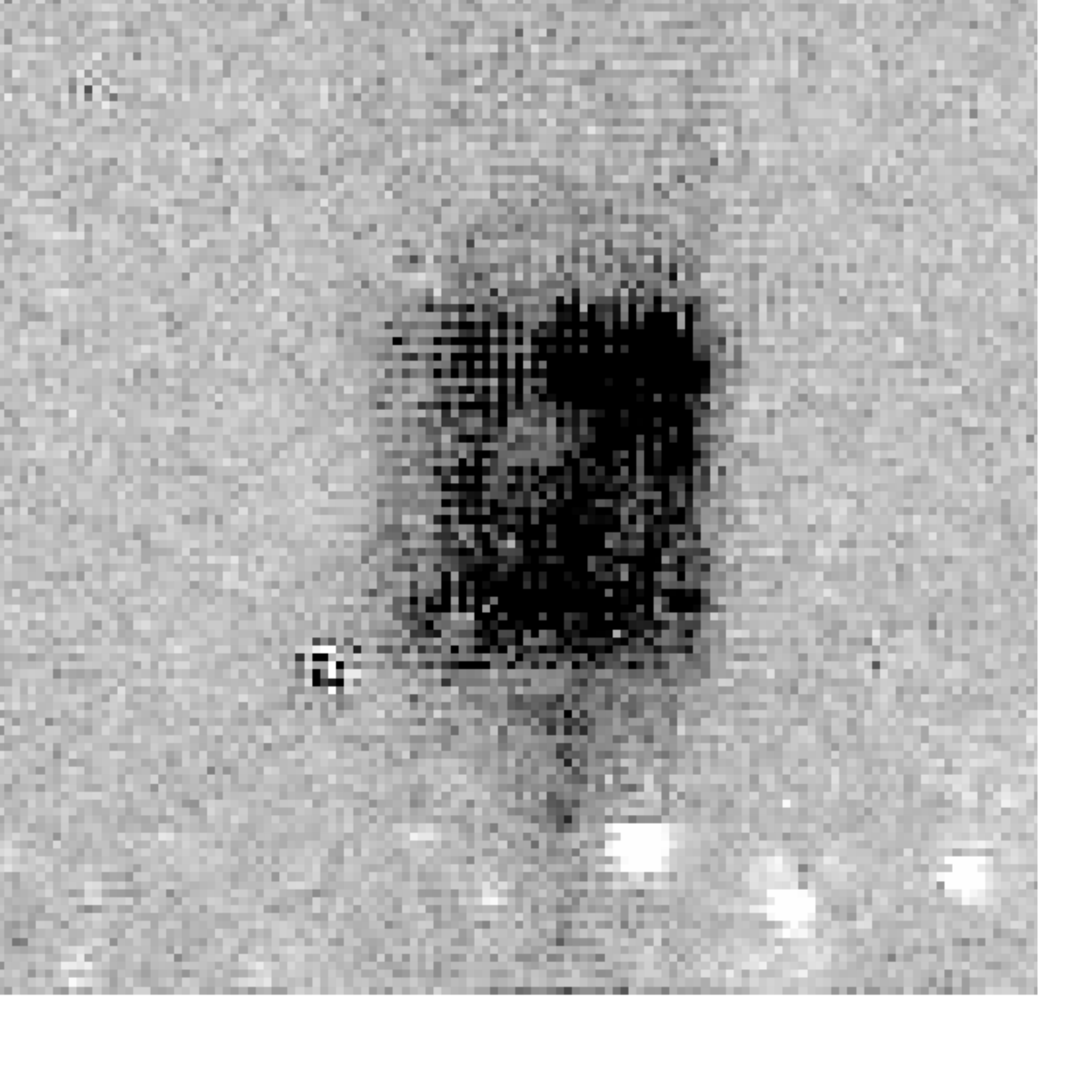} \hfill
\includegraphics[width=2.75cm,angle=0]{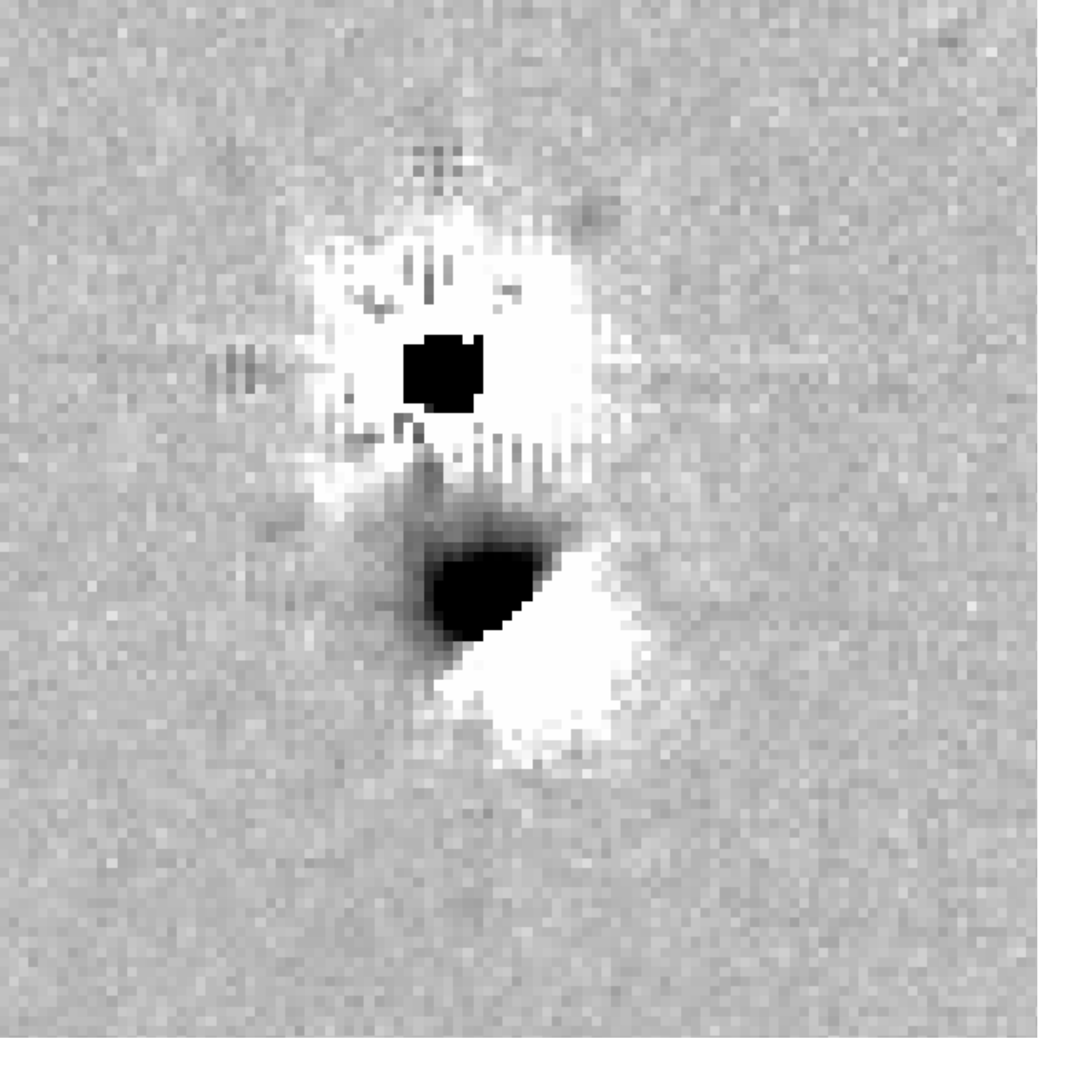} \\

\caption{\label{artifacts} Example of false positives in the \htwo$-$K
difference images. Suspected \htwo\ bright features are darker. From the top
left to the bottom right, the panels show the following: i) electronic
cross-talk from bright stars; ii) reflections from very bright stars; iii)
diffraction rings around bright stars; iv) a variable (brighter during \htwo\
imaging) and saturated star; v) reflection from bright star just outside the
edge (at bottom) of an image; vi) high proper motion star (next to variable star
-- fainter during \htwo\ imaging.}

\end{figure}

These \htwo$-$K difference images show many real \htwo\  emission line objects
(such as shock excited Jets, SNRs or PNe), but also a large number of \htwo\ 
false positives caused by image and data analysis artifacts, as well as
variability. In Fig.\,\ref{artifacts} we show six examples of such potential
false positives. Stars which are saturated or near the saturation limit (of
about K=11\,mag) are not subtracted completely. Furthermore, the K-band and
\htwo\ images are typically taken several years apart from each other. Hence,
any object that is variable, such as some giant stars or YSOs, might leave a
positive or negative residual in the difference images, as do high proper motion
stars. Furthermore, several kinds of image artifacts, such as reflections,
memory effects from bright stars, and electronic cross-talk also leave residuals
resembling \htwo\ emission line features. Finally, in cases where the two images
had very different seeing, most stars were not removed completely.

\subsubsection{Extended Source Detection}

Most of the real  \htwo\ features in our images are spatially resolved and have
a low surface brightness. Furthermore, many of the extended features are
projected onto a spatially variable background. Hence, before the detection of
these extended, low surface brightness features, we filtered our images to
remove remaining point sources and large scale variable background. We replace
any small-scale structures (less than 2\arcsec\,x\,2\arcsec) that have a pixel
value exceeding the 5$\sigma$ noise in the images with the local background.
This will remove most un-subtracted point sources. We determined the {\it local}
background as the median pixel value within 20\arcsec\ and subtracted it from
the \htwo$-$K difference images. Note that this will remove some of the largest
scale features such as extended \hii\ regions from the catalogue. However, in
most cases a significant fraction of this emission will still be detected as
several individual, smaller features. Hence in general we will have some
detections of most extended objects. Readers interested in particular, very
extended objects should however, re-process our \htwo$-$K difference images with
an appropriate spatial filter. 

We identified every region in the background subtracted and point source removed
images which was larger than four square arcseconds and had pixel values above
half the {\it rms} noise in the images. This was done by plotting contours in
ds9\footnote{http://ds9.si.edu/site/Home.html} at the respective level. The
shape of each closed contour is described by a polygon and is referred to as a
{\it 'region'} hereafter. The minimum size limit is essential to remove most of
the remaining point sources and noise from the list of objects. We rejected
every region that had a 2MASS point source within three arcseconds from the
region centre to also automatically remove the majority of saturated stars from
our list. Furthermore, many very bright stars (K\,$<$\,7\,mag) showed
diffraction rings (e.g. top right panel in Fig.\,\ref{artifacts}) that our
procedure would pick up. We thus also removed every region that was completely
within 35\arcsec\ (slightly larger than the radius of the diffraction rings)
from one of these very bright stars. Finally, all regions within 10\arcsec\ from
the edge of an image are removed. Note that the overlap between images is
generally larger than this, hence no objects are lost in gaps. There is a small
number of objects which have indeed multiple entries in the catalogue as they
are detected (in whole or in part) on more than one image. We have not removed
or joined these multiple entries in the final catalogue.

The requirements for the automated source detection (4 square arcseconds above
the 0.5\,$\sigma$ single pixel noise level), combined with the
0\fasec.2\,$\times$\,0\fasec.2 pixel size, can be used to estimate the
detection limit. In essence the software will pick out any extended object
whose surface brightness is higher than the 5\,$\sigma$ one pixel noise listed
for every image. The one pixel noise values for all images are listed
Table\,\ref{imagedatatable} in the Appendix. 

\begin{figure}
\includegraphics[width=4.19cm,angle=0]{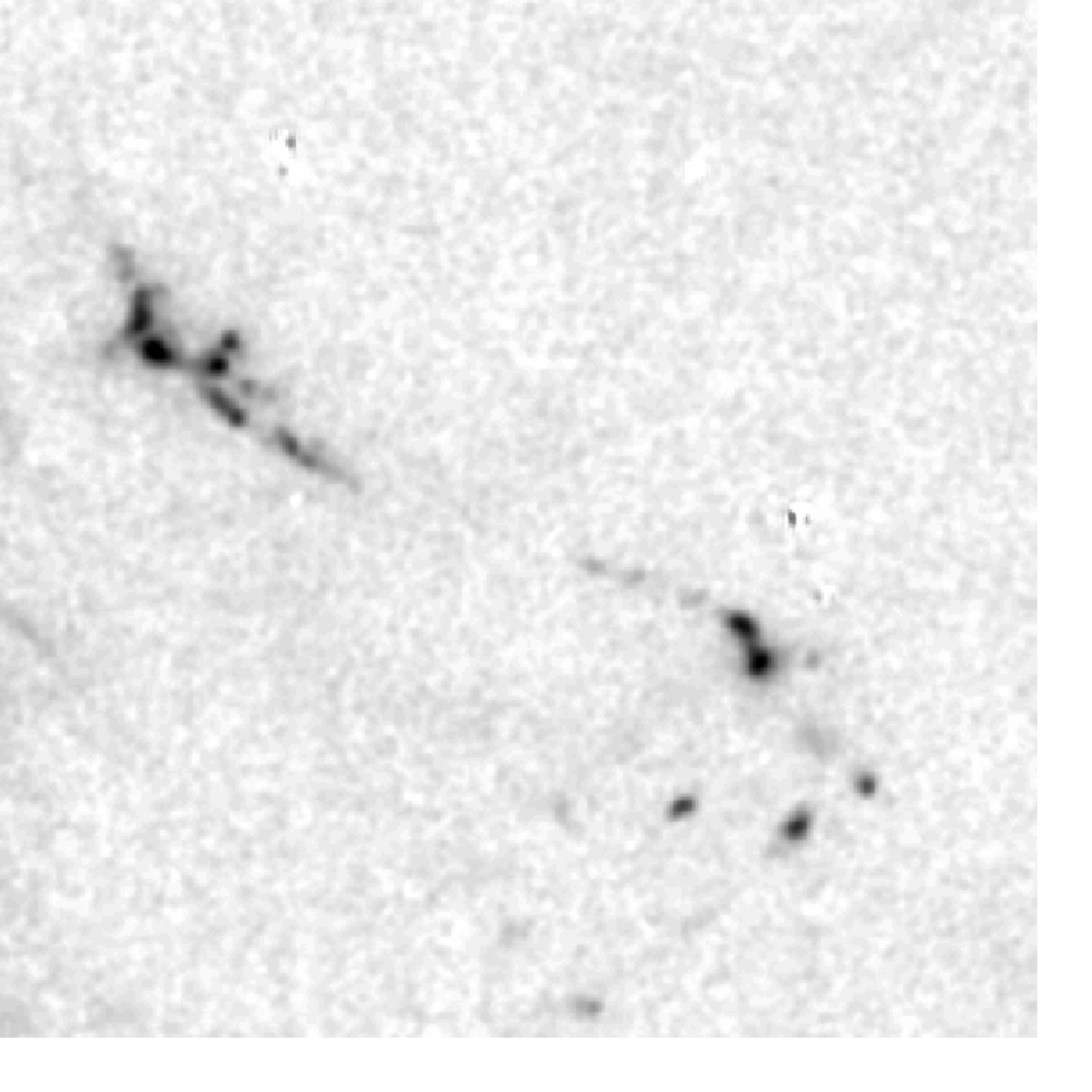} \hfill
\includegraphics[width=4.19cm,angle=0]{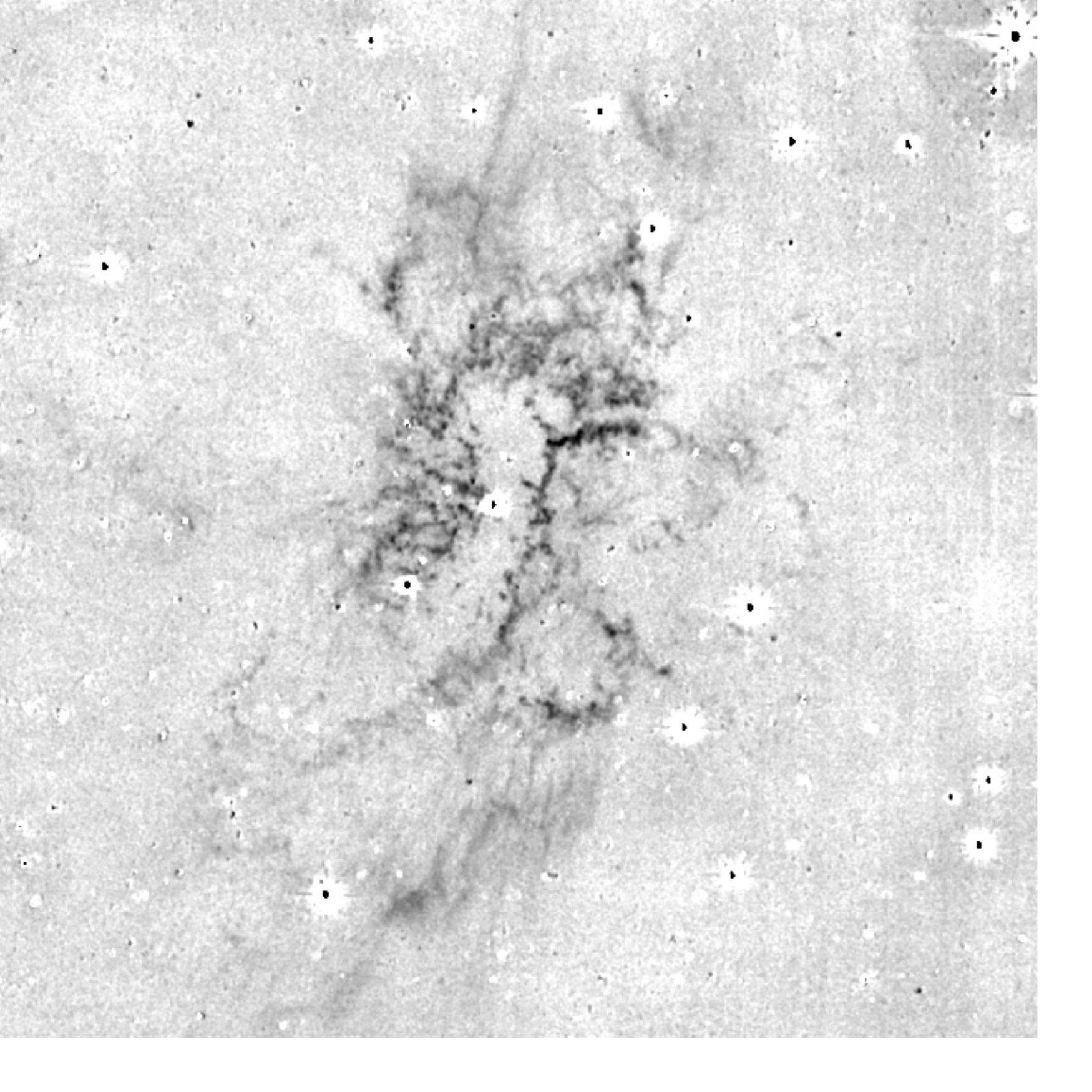} \\
\includegraphics[width=4.19cm,angle=0]{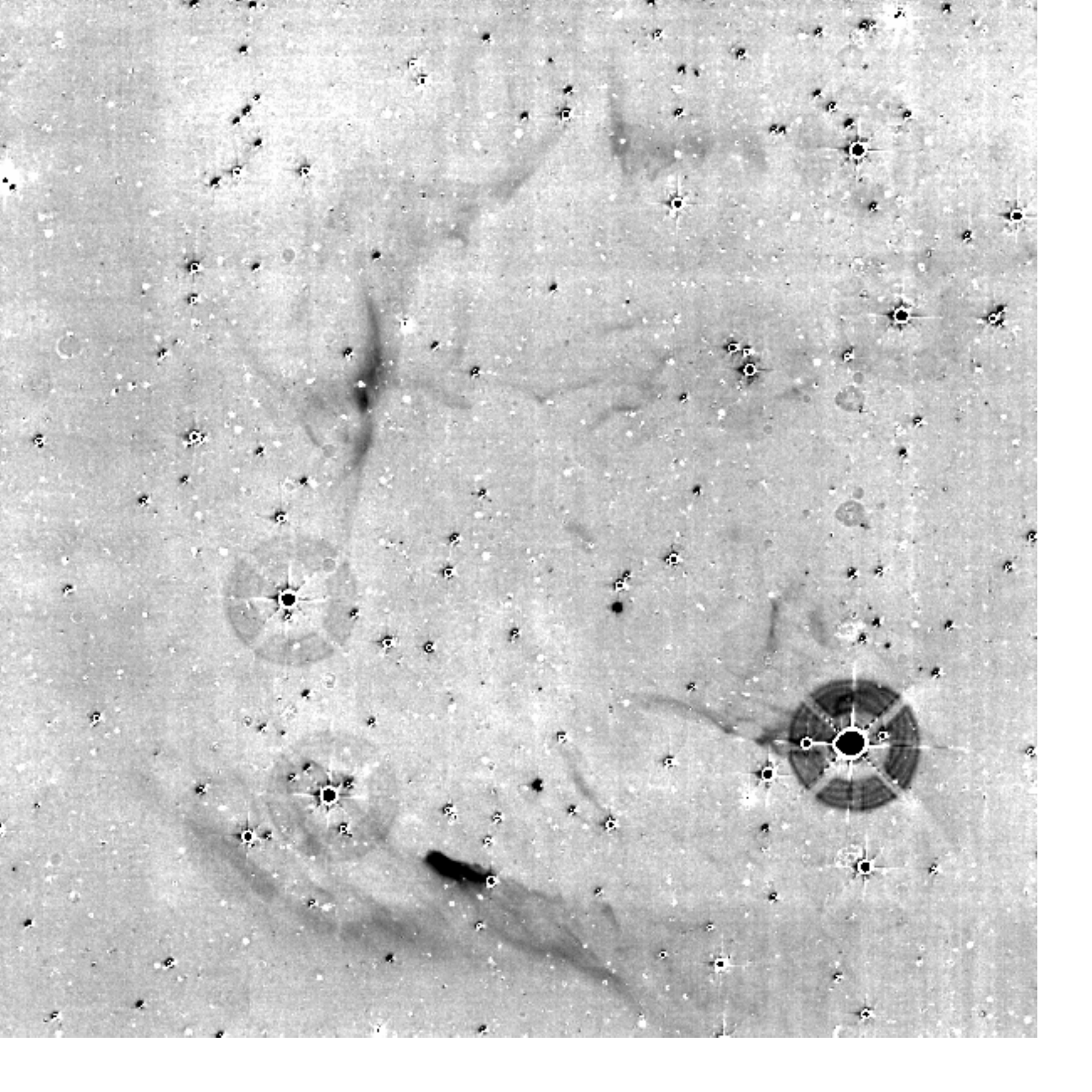} \hfill
\includegraphics[width=4.19cm,angle=0]{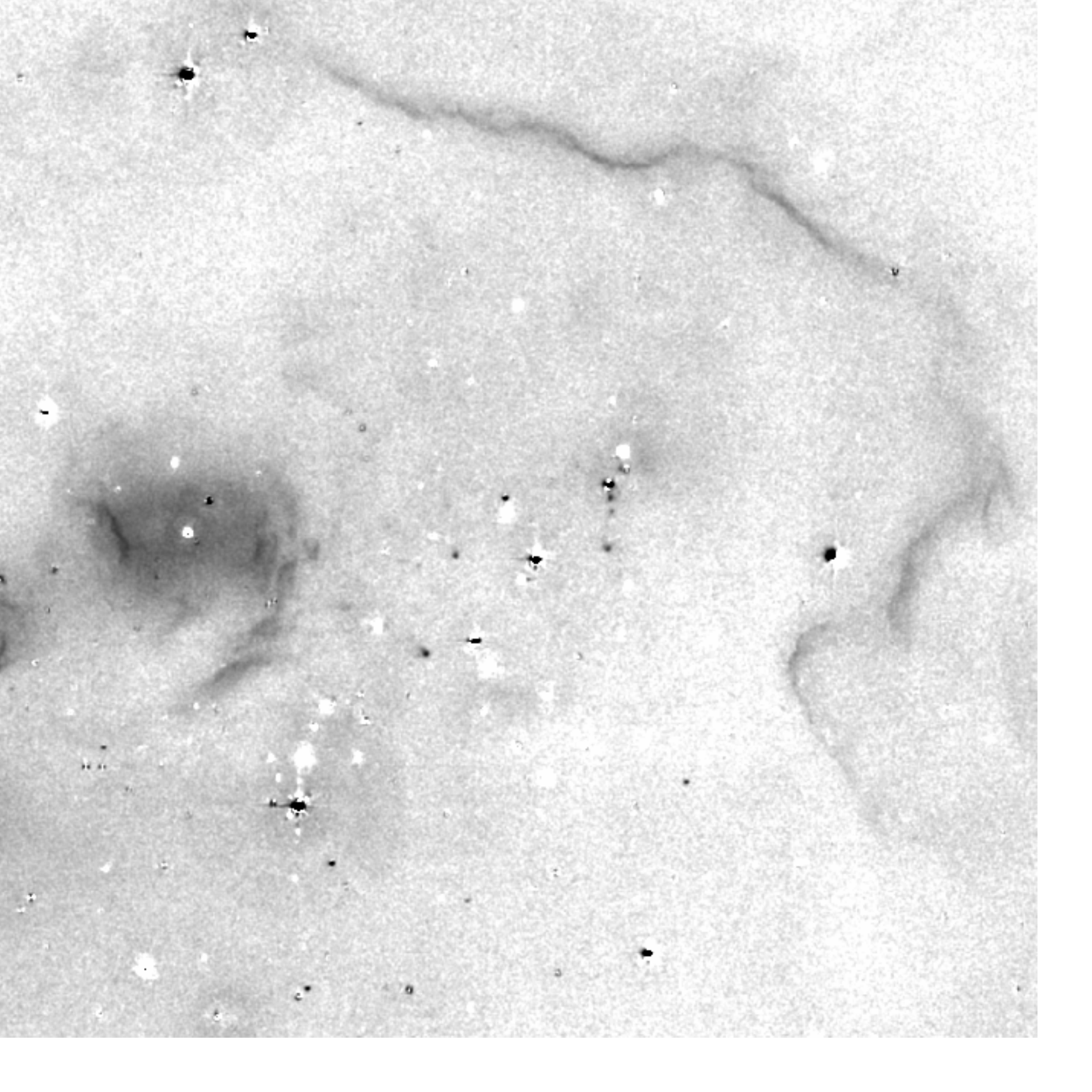} \\

\caption{\label{realobjects} Example of real objects in the \htwo$-$K difference
images for each of our object categories. Suspected \htwo--bright features are
darker. From the top left to the bottom right, the panels show the following: i)
'j' -- outflow from IRAS\,20294+4255; ii) 'p' -- Planetary Nebula SH\,2-71; iii)
's' -- Supernova remnant G11.2$-$0.3; iv) 'u' -- Emission near the cluster
VDB\,130.}

\end{figure}

\subsubsection{Source verification/classification}

The above automated detection procedure still included a large number of regions
which were obviously not real \htwo\ features, and removed others which happened
to be in the vicinity of bright stars. Hence, we manually checked all images to
remove any region that was obviously not a real \htwo\ feature, e.g. image
artifacts, variable or saturated stars and to re-add regions that were removed
but clearly real. About 35\,\% of all images were searched by two people
independently to gain an understanding of the completeness and contamination of
the selected \htwo\ features. The remaining images were only searched by one
person. Based on the comparison of the catalogues obtained for the images with
two people selecting objects as real, we estimate that the contamination of the
catalogue with image artifacts or noise is very small. At most 1\,--\,2\,\% of
the catalogue entries might be artifacts. Also the completeness of the
catalogues is very high. We estimate that more than 95\,\% of all the real
\htwo\ features detected automatically are in the final catalogue. Missing
objects are usually small features in regions of large extended \htwo\ emission.
These missing objects do not contribute with any significance to the total area
or flux of the \htwo\ emission line catalogue. 

During the above discussed manual verification of the automatically selected
\htwo\ features, we also classified each feature into one of four categories: i)
'j' for all objects which seem to be part of a Jet or outflow from a young star.
This classification was based on the shape of the feature, as well as its
appearance/colour in the JK\htwo\ colour images. E.g. isolated, extended, high
surface brightness \htwo\ knots which are situated in or near obvious star
forming regions are classified as jet/outflow; ii) 'p' for objects that resemble
known PNe. These tend to be ring-like, bipolar or in some cases more complex
structures, typically not related to star forming regions. Note that in some
cases the appearance alone is not sufficient to distinguish a bipolar PN from a
Jet emanating from a young star. We furthermore checked the positions of all
known PNe and candidates in the survey area and classified all \htwo\ features
as 'p' if they were within a few arcsecond of a known object. We utilised the
PNe entries in SIMBAD and the catalogues from IPHAS \citep{2014MNRAS.443.3388S},
MASH \citep{2006MNRAS.373...79P} and MASH2 \citep{2008MNRAS.384..525M}; iii) 's'
for objects which are most likely part of a Galactic SNR. All \htwo\ features
within the area of a known SNR (we utilised the list of
\citet{2009BASI...37...45G}) were selected to be of this category if they were
not part of an obvious PN or Jet/outflow, or a small individual feature with no
resemblance of the \htwo\ emission in other SNRs; iv) 'u' for all objects which
could not be assigned to any of the other three categories. Note that the vast
majority of these features are most likely part of PDRs surrounding \hii\
regions. Thus, we refer to all the unclassified regions as PDRs. In
Fig.\,\ref{realobjects} we show one example of each of the object categories. 

Note that the source selection and classifications (except for the SNRs and the
known PNe) were done blind, i.e. without using any catalogues of known objects
or SIMBAD.  This hence gives us a further estimate of the completeness and
accuracy of the classification by comparing to lists of known objects. We
utilised the catalogue of Molecular Hydrogen emission line Objects (MHOs) from
\citet{2012MNRAS.421.3257I} who manually searched about 33 square degrees of
early UWISH2 data for emission from jets and outflows. They list 134 MHOs and we
have checked what fraction of these are contained in our catalogue: 83\,\% of
the MHOs are included in our extended--\htwo\ feature catalogue. Exclusively all
of the non-detections (17\,\%) are faint and small \htwo\ features which in most
cases are similar to variable point sources rather than \htwo\ emission line
objects. Of the detected \htwo\ features, 79\,\% are also classified as being
part of a jet or outflow, 15\,\% are not classified ('u'), 4\,\% coincide with
emission from SNRs and 2\,\% (2 MHOs) are listed as PN candidates in our list.
Hence any objects missing in our catalogue are most likely faint and compact --
indistinguishable from variable point sources. 

\subsection{Photometry}\label{calibration}

\subsubsection*{Flux measurement and calibration}

Photometry has been obtained for each region in the \htwo$-$K images. As these
difference images are obtained by only scaling the K-band continuum fluxes, the
\htwo\ flux in all the images is conserved. We identify all pixels inside each
region and determine their median, maximum and total number of counts. We then
correct these values by the local background counts. These are estimated as the
median count value in a ring around each region with an inner radius equal to
the radius of the region and an outer radius of twice this. Note that in some
rare cases, this background estimate will be wrong, e.g. if a small region is
situated close to a larger region of extended \htwo\ emission. These occurrences
are rare, but might lead to background corrected fluxes which are erroneous or
even negative. For a further discussion of uncertainties of the photometry see
the end of this Section.

We convert the counts in each region into fluxes or surface brightness in two
steps. Firstly the counts are converted into a magnitude via:

\begin{equation}
m = m_{\tt zp} - 0.05 \cdot \left( X - 1 \right) - 2.5 \cdot \log_{10} \left(
\frac{\tt counts}{t_{exp}} \right) - m_{\tt ap}
\end{equation}
where $m_{\tt zp}$ is the magnitude zero point for the observations, $X$ the
airmass during the observations, $t_{exp}$ the exposure time in seconds and
$m_{\tt ap}$ the aperture correction. As we are only considering extended
sources we can set the aperture correction term to zero. All other terms are
obtained from the FITS header in the \htwo\ images. Note that all our
observations are taken with 720\,s integration time per pixel and the airmass is
always between one and two. Hence, in most cases, the airmass term is of the
same order or smaller than the uncertainty in $m_{\tt zp}$. Furthermore, as can
be seen in the right panel of Fig.\,\ref{seeinghist}, the general variations in
$m_{\tt zp}$ are also only of the same order of magnitude. Note that $m_{\tt
zp}$ includes the corrections that need to be made caused by the micro-stepping
and hence 0\fasec.2 pixel size in our data. 

These magnitudes are converted into fluxes by:

\begin{equation}
F = F_0^{H_2} \cdot 10^{ - \frac{m}{2.5}}
\end{equation}
The $m_{\tt zp}$ values will calibrate the magnitudes into the 2MASS K-band. We
thus used the 2MASS K-band flux zero point of
4.283$\cdot$10$^{-10}$\,W\,m$^{-2}$\,$\mu$m$^{-1}$ from
\citet{2003AJ....126.1090C} and the K-band filter width of 0.262\,$\mu$m to
determine the flux zero point $F_0^{H_2} = 1.12 \cdot 10^{-10}$\,W\,m$^{-2}$. In
combination the flux $F$ corresponding to {\tt counts} is determined as:

\begin{equation}
F = {\tt counts} \cdot \frac{1.12 \cdot 10^{-10} W m^{-2}}{t_{exp}[s]} \cdot
10^{\left\{ - \frac{m_{\tt zp} - 0.05 \cdot \left( X - 1 \right)}{2.5}\right\}}
\end{equation}

We compared our calibrated flux values to published flux values of jet knot and
SNRs and they agree at the 10\,\% level. Conversion of fluxes into surface
brightness is done using the pixel size of 0.04 square arcseconds in our
images. 

\begin{figure}
\includegraphics[width=8.6cm,angle=0]{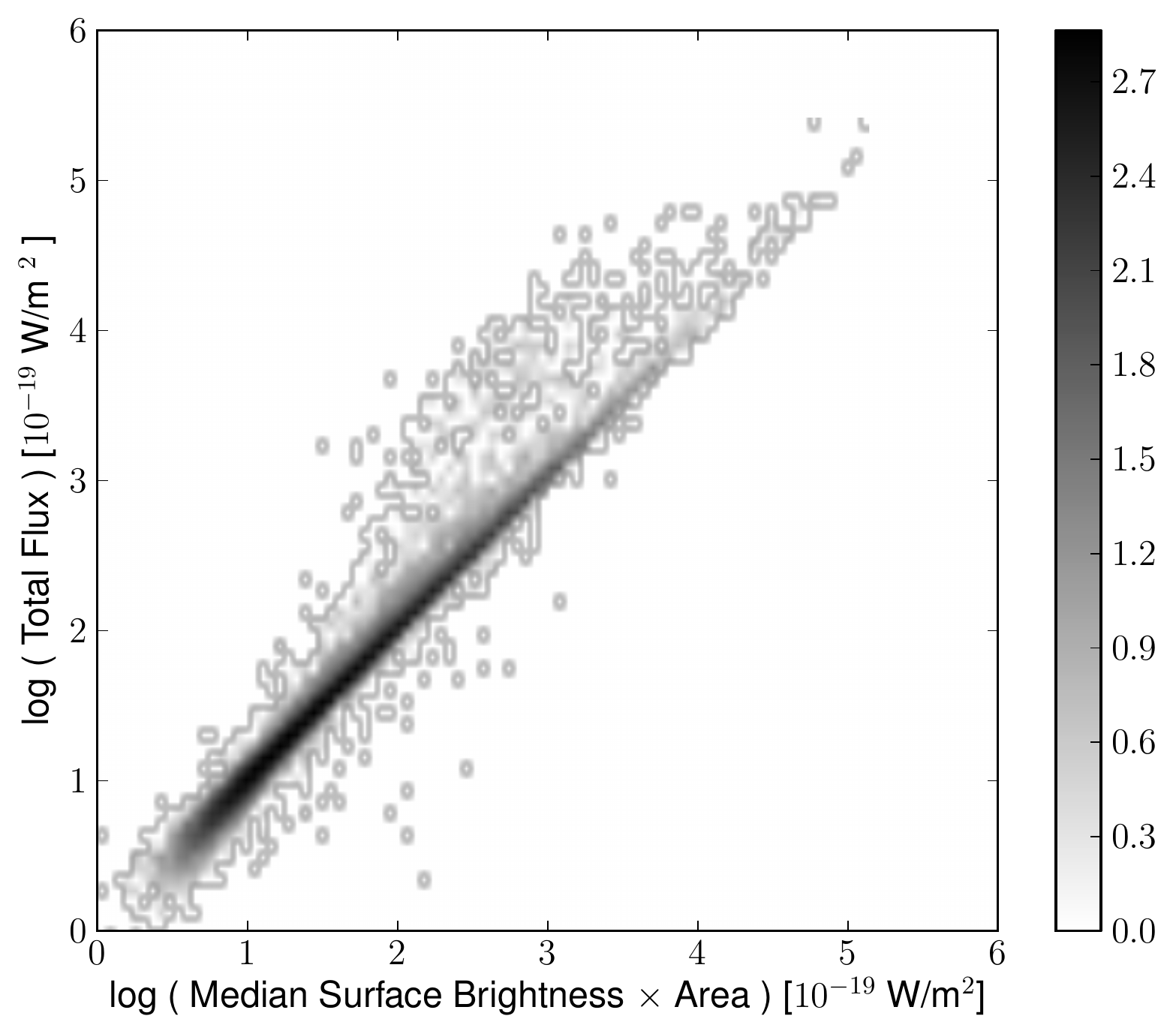} \\
\includegraphics[width=8.6cm,angle=0]{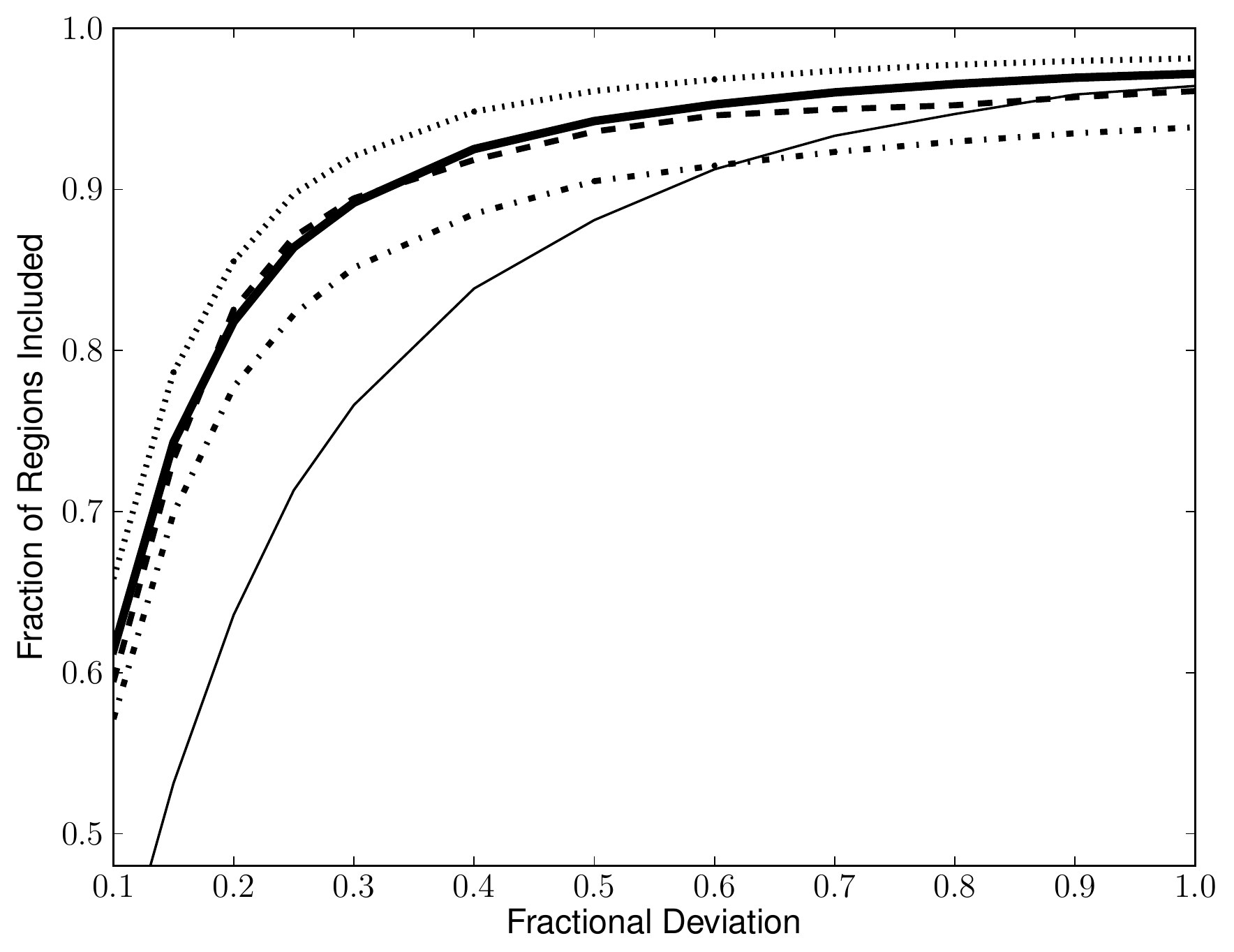}

\caption{\label{totalflux} Top: Comparison of total flux and median surface
brightness times area for all detected objects. The gray scale indicates the
Log of the density of objects in each position. Most objects are very close to
the 1:1 line. Bottom: Fractional deviation of the two flux estimates from the
top panel vs. the fraction of objects which have a deviation smaller than it.
The different line styles indicate the various objects: dot-dash line for SNRs,
dashed line for PNe, bold solid line for all objects, dotted line for PDRs,
thin solid line for jet features.}

\end{figure}

\subsubsection*{Total flux estimates for objects}

There are two ways to determine the entire \htwo\ flux of a region. We can: i)
use the background corrected {\it total flux} inside a region ($F_{tot}$), or we
can ii) multiply the background corrected median surface brightness by the area
(or {\it median flux}) of the region ($F_{med}$). Both ways have obvious
drawbacks. If the region contains a non-subtracted residual, either due to a
saturation or variability (see also Fig.\,\ref{artifacts}), then the total flux
can be influenced (positively or negatively) by the presence of the star. If the
median surface brightness is not a good estimate of the average \htwo\ flux,
e.g. when a small fraction of the region contains a large fraction of the flux,
the total flux will be underestimated. In order to compare the two methods we
compare the estimated fluxes for both in the top panel of Fig.\,\ref{totalflux}.
It shows the density of objects, and in the majority of cases the two \htwo\
flux estimates are comparable. There are, however, a number of cases where the
two estimates disagree by a large amount. 

We investigate for what fraction of objects the two flux estimates agree within
a given range. This is shown in the bottom panel of Fig.\,\ref{totalflux}. The
x-axis in the plot is the fractional deviation of the two fluxes (from 10\,\%
to a factor of two) and the y-axis shows the fraction of objects with a
deviation smaller than this. The different line styles indicate the various
object types. The bold solid line is for all objects, the thin solid line for
jet features, the dotted line for PDRs, the dashed line for PNe and the
dot-dash line for SNRs. As one can see from the figure, for about 80\,\% of all
objects the deviation of the two flux estimates is less than 20\,\%. For the
jet features the agreement is clearly worse, with only about 65\,\% having flux
estimates with a better than 20\,\% agreement.

The reason for the latter seems to be the surface brightness distribution
within jet knots. Most of them contain a small number of pixels which
contribute a large fraction of the flux. This seems to be less an issue for
PNe, PDRs and SNRs. We hence recommend to use $F_{tot}$ for the total fluxes of
all objects classified as jet, while for the other object types $F_{med}$ seems
more appropriate, as it prevents the potential inclusion of flux from
(unsubstracted or variable) stars projected onto the \htwo\ emission line
object. However, should the reader require more accurate photometry of selected
objects, we recommend that they redo the flux estimate in our images, ensuring
that only \htwo\ emitting areas are included in the photometry.

\subsection{Object groups}

Many of the detected \htwo\ emission features are not isolated, but are rather
part of a group of objects. This is particularly true for the PNe and SNRs, 
which often consist of several emission features due to low surface brightness
or large extent. We have hence grouped regions according to their spatial
distribution. Objects were considered part of a group if they had a nearest
neighbour within a given angular distance. For each group we determined
properties such as the position, size (as the radius of a circle enclosing all
group members) and total flux.

In the case of PNe regions, these groups can be considered as actual PN. We
grouped objects automatically if they were separated by less than 3
arcminutes. Additionally we inspected all of these PNe visually to ensure that
there were no two PNe closer to each other than the 3' threshold and that the
extended PNe in the catalogue had no 'outlying' features that were classified
as a separate PN.

For the Jets and outflows it is not a simple task to identify which jet knots
are part of which outflow and which object is the actual driving source of the
jet. Such a procedure needs a detailed study of each region and is beyond the
scope of this paper. However, groups of Jet/outflow knots can be considered as
star forming regions with actively accreting YSOs. Given that a typical distance
of jets and outflows in the survey area is about 3.5\,kpc
\citep{2012MNRAS.425.1380I} and the jets seem to occur in small groups of about
5\,pc size \citep{2012MNRAS.421.3257I}, we used 0\fdeg.1 as minimum distance to
separate groups of jets and outflows. Hence, these groups can be viewed as very
young active star forming regions, slightly more extended than a typical young
cluster (e.g. \cite{2008MNRAS.389.1209S}). Note that if moved to 3\,kpc, the
jets and outflows in NGC\,1333 would be distributed over an area of about
2'\,x\,2' on the sky.

The objects classified as 'u' (most likely \hii\ regions or PDRs) are also
grouped with the same minimum distance of 0\fdeg.1, as they are probably at the
same typical distances as the jet and outflow features. In essence these groups
are likely to represent more evolved regions of star formation where the \htwo\ 
emission is caused by sources of ionising radiation.

We do not group the SNR objects in the same way as the other object types, since
many of the SNRs are very extended on the sky. Instead, we have manually
selected all the \htwo\ features which are part of each of the identified SNRs. 

\subsection{Catalogue description}
 
The full extended \htwo\ feature catalogue displayed in
Table\,\ref{objectdatatable} contains the following columns:

\begin{enumerate}

\item Object ID; this is derived from the Galactic coordinates of the centre of
each region. As centre we use the geometric centre of the polygon enclosing the
detected \htwo\ emission.

\item Right Ascension and Declination (J2000) of the centre of the emission
region.

\item Area $A$ of the emission region in square arcseconds.

\item Radius $r$ of the emission region in arcseconds; This is the minimum radius
of a circle around the centre of the region that is enclosing all the emission.

\item  Median surface brightness F$_{sb}^{med}$ of the region in
10$^{-19}$\,W\,m$^{-2}$\,arcsec$^{-2}$; This is the surface brightness
determined from the background corrected median intensity in the region. 

 %
 %

\item Peak surface brightness F$_{sb}^{max}$ of the region in
10$^{-19}$\,W\,m$^{-2}$\,arcsec$^{-2}$; This is the peak surface brightness of
the region. It might be influenced by the presence of stars inside the region.

\item One pixel {\it rms} noise surface brightness F$_\sigma$ in
10$^{-19}$\,W\,m$^{-2}$\,arcsec$^{-2}$; This is the one sigma {\it rms} of the
background in a ring with inner radius $r$ and outer radius $2r$ around each
region (determined after sigma clipping to remove remaining stars and real
emission features). 

 %
 %

\item Total brightness F$^{tot}$of the region in 10$^{-19}$\,W\,m$^{-2}$; This
is the total flux measured in each region. It might be influenced by the
presence of stars inside the region. An alternative measure of the total flux
would be the product of the median surface brightness and area of the region.

\item Relative uncertainty $\Delta F / F$, in percent, of all fluxes due to the
uncertainty in the magnitude zero point of the observations $\Delta m_{\tt zp}$.

\item Classification $C$ of the object; This is a letter indicating what kind
of object the region is most likely a part of. These are: j -- jet or outflow
from a YSO; p -- part of a PN; s -- part of a SNR; u -- unknown nature, most
likely part of a PDR near an \hii\ region.

\item Name of the tile the region is on.

\item Name of the image the region is on.

\item Group identifier the object belongs to. The group identifier contains the
object type, as well as the Galactic coordinates of the group, calculated as the
geometric centre of the features that make up the group.

\end{enumerate}


\section{Results and Discussion}\label{discussion}

In this paper we will only discuss the general distribution and properties of
the detected \htwo\ emission regions. For a detailed discussion of individual
objects, or groups of objects we refer the reader to publications in
preparation.

\begin{table*}

\caption{\label{regionresulttable} Table showing the different parts of the
survey used in the analysis and some of the accumulated properties of the
objects identified in them. We list the survey area, the sum of the total fluxes
for each of the four object types and the total number of identified groups of
objects. Note that for PNe and SNRs a 'group' identifies individual objects,
while for Jets and PDRs a group simply refers to a group of spatially related
individual \htwo\ features. The numbers in brackets indicate the fraction (as
percent, rounded to the nearest integer) of the total flux or numbers of the
total in the entire survey. We separate inner and outer GP at $l =$\,30\degr\
and list all values for both together in the additional row labeled 'GP'.}

\begin{center}
\begin{tabular}{lcccccccccccc}
\hline
Region & Area & & F$_{\rm tot}^{\rm Jet}$ & F$_{\rm tot}^{\rm PDR}$ & F$_{\rm tot}^{\rm PN}$ & F$_{\rm tot}^{\rm SNR}$ & & N$^{\rm Jet}$ & N$^{\rm PDR}$  & N$^{\rm PN}$ & N$^{\rm SNR}$  \\
  &  [deg$^2$ (\%)] & & \multicolumn{4}{c}{[10$^{-14}$\,W\,m$^{-2}$] (\%)} & & \multicolumn{4}{c}{[Number (\%)]}  \\
\hline
Total                & 286.45 & & 10.6 & 49.6 & 7.67 & 46.1 & & 711 & 1309 & 284 & 30 \\
\hline
GP                  & 209.00 (73) & & 4.70 (44) & 30.3 (61) & 7.37 (96) & 46.1 (100) & & 450 (63) & 925 (71) &  261 (92) & 30 (100) \\
\hline
GP\,(Inner)  & 95.18 (33) & & 2.83 (27)  & 24.7 (50) & 5.73 (75) & 15.5 (34) & & 253 (36) & 489 (37) & 112 (39) & 20 (67) \\
GP\,(Outer) & 113.79 (40) & & 1.86 (18) & 5.58 (11) & 1.64 (21) & 30.6 (66) & & 197 (28) & 436 (33) & 149 (52) & 10 (33) \\
Cygnus           &  41.99 (15) & & 5.73 (54) & 18.3 (37) &  0.24 (3) &  ---  & & 210 (30) & 353 (27) &  16 (6) & ---  \\
Auriga            &  35.46 (12) & & 0.12 (1) & 1.13 (2) &  0.055 (1) &  ---  & &  51 (7) &  31 (2) &  7 (2) & --- \\
\hline
\end{tabular}
\end{center}
\end{table*}

\begin{table*}

\caption{\label{regionaverages} Table listing the density (G, in objects per
square degree) of the groups of objects in the various parts of the survey. We
also list the median total flux ($\bar{\rm F}_{\rm tot}$ in
10$^{-19}$\,W\,m$^{-2}$) for each kind of group of objects. Note that for PNe
and SNRs a 'group' identifies individual objects, while for Jets and PDRs a
group simply refers to a group of spatially related individual \htwo\ features. 
We separate inner and outer GP at $l =$\,30\degr\ and list all values for both
together in the additional row labeled 'GP'.}

\renewcommand{\tabcolsep}{5pt}
\begin{center}
\begin{tabular}{lcccccccccc}
\hline
Region &   & G$^{\rm Jet}$ & G$^{\rm PDR}$ & G$^{\rm PN}$ & G$^{\rm SNR}$ & & $\bar{\rm F}_{\rm tot}^{\rm Jet}$ & $\bar{\rm F}_{\rm tot}^{\rm PDR}$ & $\bar{\rm F}_{\rm tot}^{\rm PN}$ & $\bar{\rm F}_{\rm tot}^{\rm SNR}$ \\ 
 & &  \multicolumn{4}{c}{ [objects deg$^{-2}$]} & & \multicolumn{4}{c}{ [10$^{-19}$\,W\,m$^{-2}$]} \\ 
\hline
Total       &        & 2.48 & 4.57 & 0.99 & 0.10 & & 193 & 149 & 441 & 7096 \\
\hline
GP        &          & 2.15 & 4.43 & 1.25 & 0.14 & & 207 & 137 & 453 & 7096 \\
\hline
GP\,(inner) & & 2.66 & 5.14 & 1.18 & 0.21 & & 298 & 228 & 570 & 4345 \\
GP\,(outer) & & 1.73 & 3.83 & 1.31 & 0.09 & & 160 & 86.1 & 366 & 18120 \\
Cygnus          & & 5.00 & 8.41 & 0.38 & ---  & & 204 & 183 & 173 & --- \\
Auriga          &  & 1.44 & 0.87 & 0.20 & ---  & & 74.5 & 203 & 376 & --- \\
\hline
\end{tabular}
\end{center}
\end{table*}

 %
 %

\subsection{General Distributions}

The entire survey region is composed of 5872 individual images. In only about
one third of them (1935) have we identified real \htwo\ emission line features.
This indicates that most areas, especially along the GP, are devoid of
detectable \htwo\ emission, and that the detected \htwo\ features are
localised/clustered. In total we detected 33200 individual extended \htwo\
emission line features. About 62\,\% of them are situated in fields along the GP
(37\,\% in the inner and 25\,\% in the outer GP -- separated at $l =$\,30\degr),
about 36\,\% are in the Cygnus area, and the remaining 2\,\% are in Auriga.
Detailed results for the identified groups of objects are outlined in
Tables\,\ref{regionresulttable} and \ref{regionaverages}.  In these tables we
break down the numbers for each of the survey regions for the different object
classes (Jets, PNe, SNRs, unclassified -- most likely PDRs). Furthermore, we
show the total area covered by each part of the survey. We list the number of
PNe, SNRs, the number of Jet groups (actively accreting star forming regions)
and groups of other \htwo\ emission features, as well as their total fluxes,
median fluxes and projected object densities in the different parts of the
survey.

\begin{figure}
\includegraphics[width=8.6cm,angle=0]{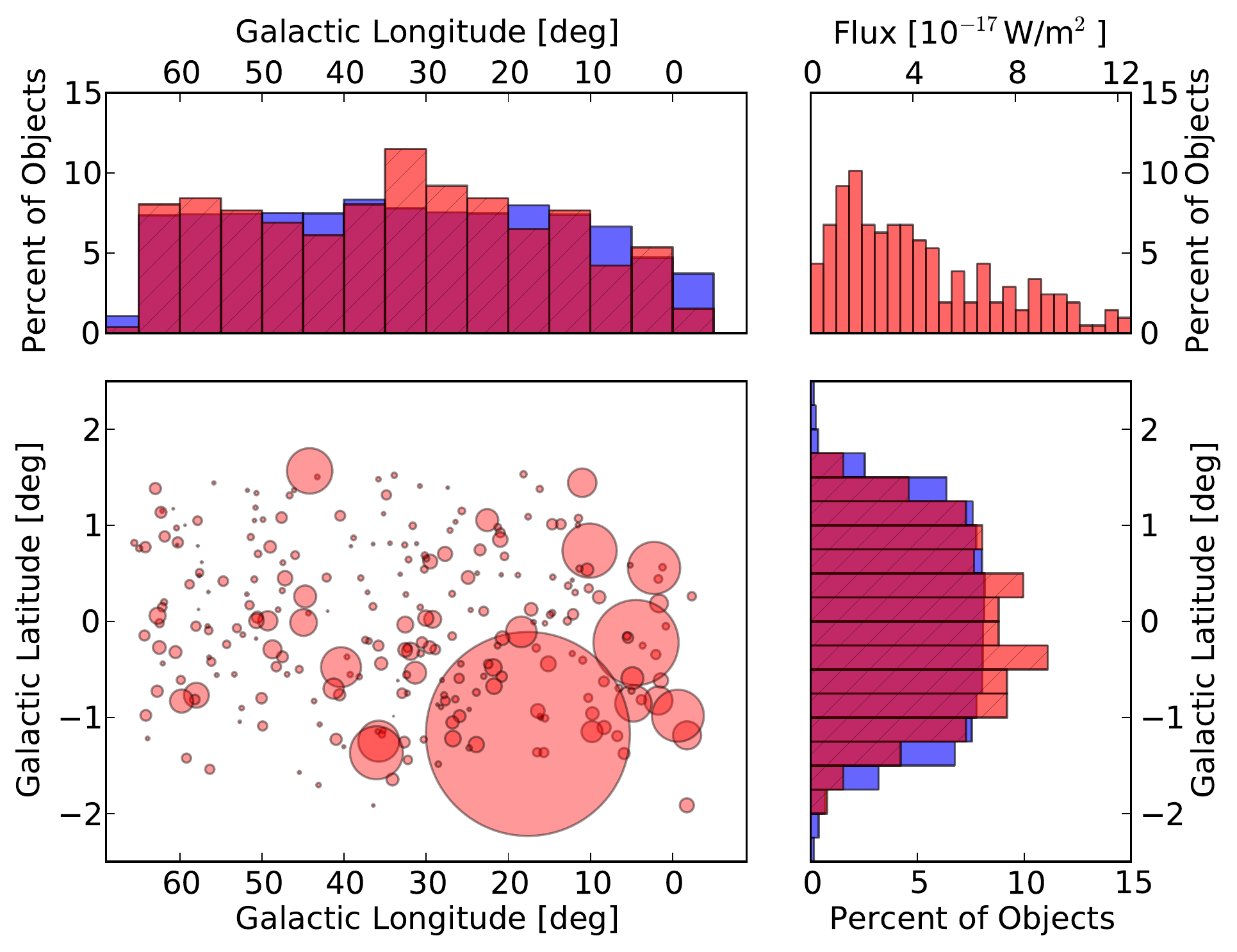} \\
\includegraphics[width=8.6cm,angle=0]{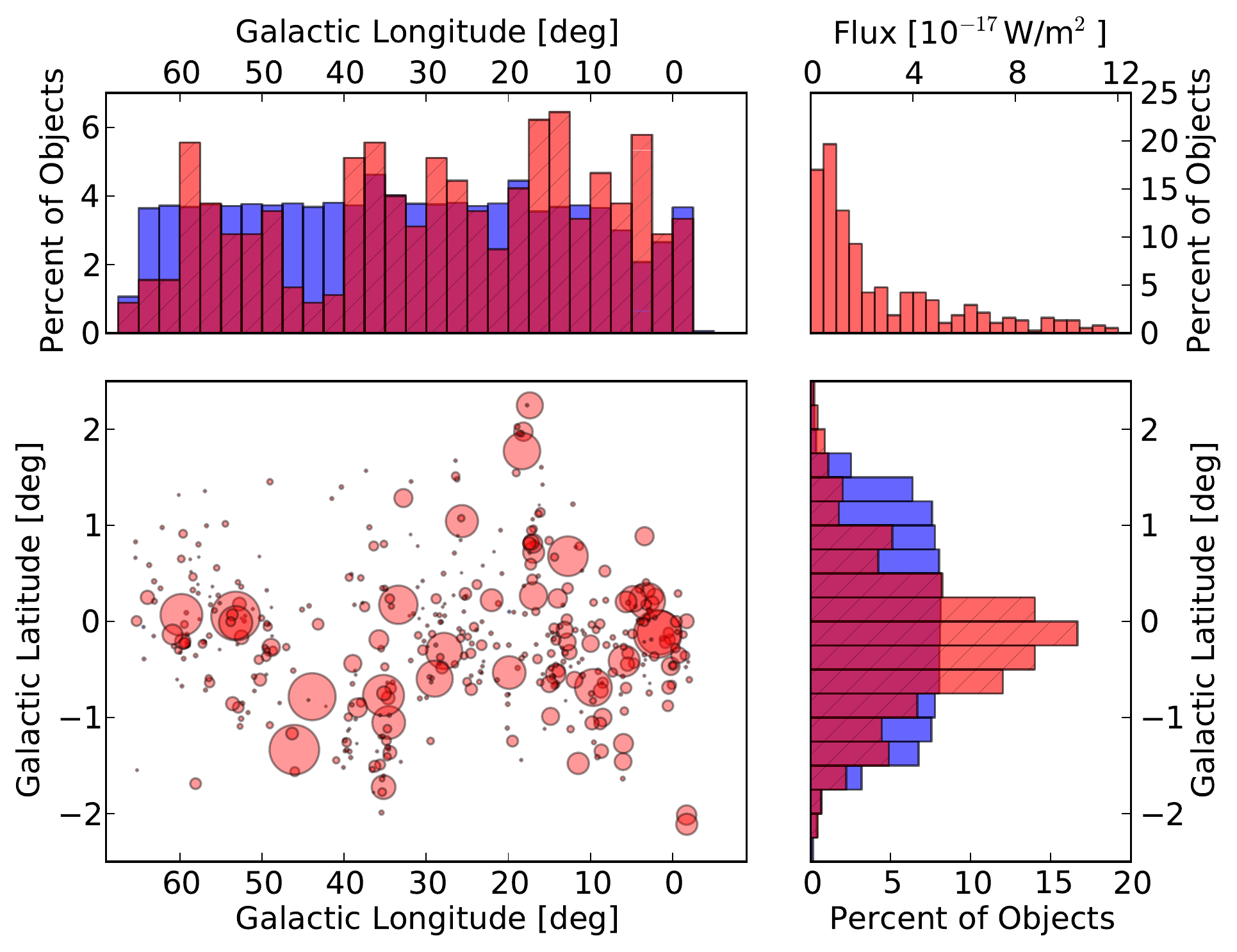} \\
\includegraphics[width=8.6cm,angle=0]{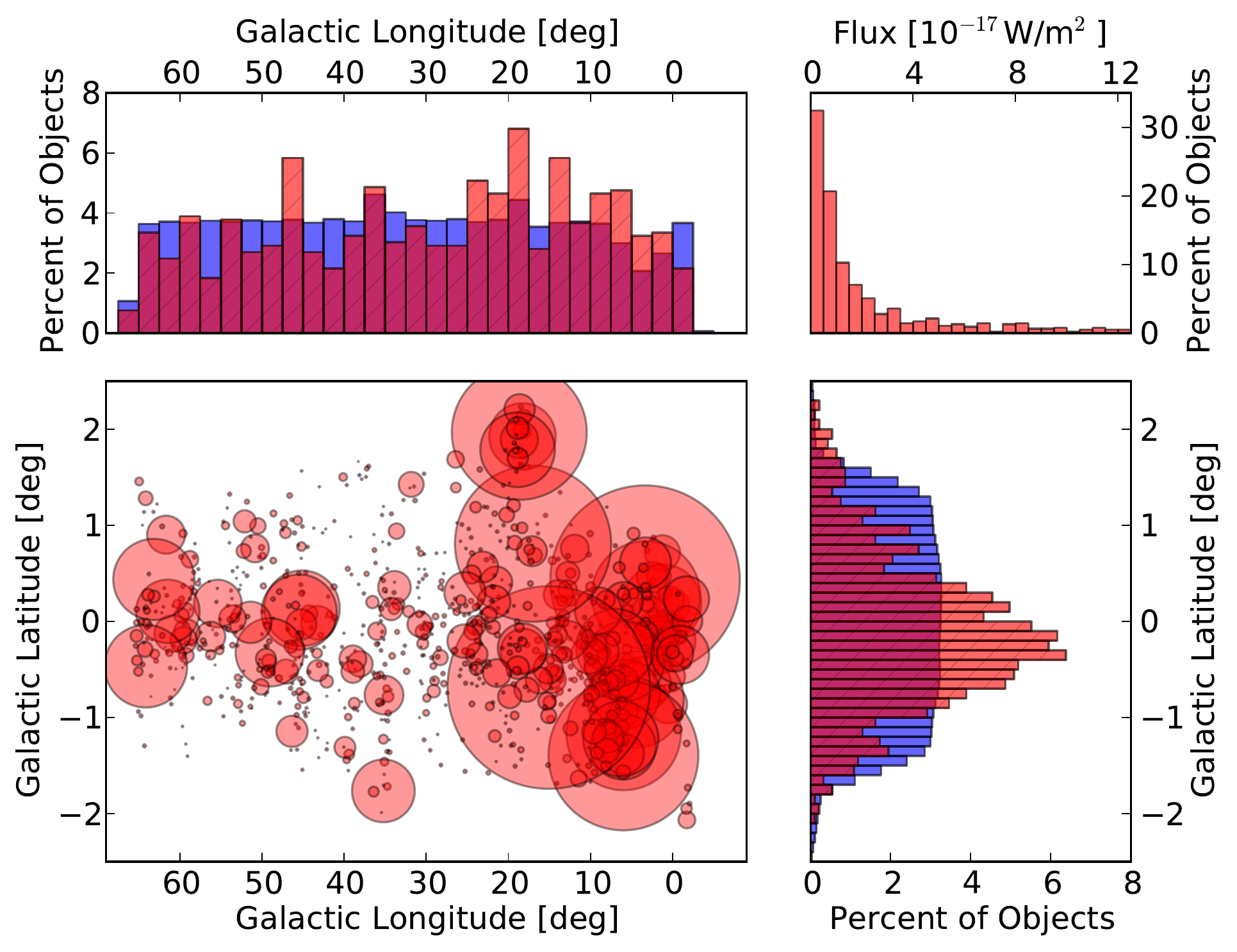} \\

\caption{\label{objdist}  Distributions of PNe (top panel),  jet groups (middle
panel) and unknown objects (mostly PDRs, bottom panel). For each object type
there are four plots. The bottom left indicates the spatial distribution where
the circle size indicates the total flux of each group. The same scaling is
applied in all plots. The top right graph show the distribution of the total
fluxes for each group. Some of the bright groups of unknown objects are beyond
the limit of the graph. The top left and bottom right graphs show the
distribution of the number of objects along and across the GP (red, hatched) vs.
the survey coverage (blue, unhatched).}  

\end{figure}

In Fig.\,\ref{objdist} we show the spatial distributions of all the groups of
objects (Jets, PNe, unknown/PDR) as well as their flux distributions. The
distributions along the GP show that the objects are distributed slightly
differently. In particular they are not in agreement with a homogeneous
distribution in our survey. The distribution indicates that there are slightly
less PNe than expected for a homogeneous distribution at Galactic longitudes
less than 20\degr--\,30\degr. This is most likely due to the higher extinction
in this direction which will lower our detection limit to smaller distances.
This is further supported by Table\,\ref{regionaverages} which indicates the
number of PNe per unit area in the inner GP is about 10\,\% lower than in the
outer GP. A KS-test shows that the PN longitude distribution has a 96.1\,\%
chance of being drawn from a homogeneous distribution. The longitude
distributions of the groups of Jets and unknown/PDR objects, both of them
representing star forming regions, are clearly different. There is a clear
overabundance of objects compared to a homogeneous distribution for longitudes
of less than 30\degr. A KS-test shows that both distributions have a probability
of only 1.4\,\% (Jets) and 3.0\,\% (unknown/PDR) to be drawn from a homogeneous
sample. It is also evident that groups (of both Jets and unknown/PDR objects)
within 30\degr\ of the GC are much brighter than the groups further away (see
Table\,\ref{regionaverages}), indicating stronger star formation activity
(traced in \htwo) closer to the GC. Furthermore, the spatial distribution of the
groups of Jets and PDRs is much more clustered than for the PNe. Hence, the
small star forming groups follow the large scale filamentary structure of GMCs
along the GP.

We also determined the scale height of the distribution of the objects
perpendicular to the GP. We utilised the method developed by
\citet{2014MNRAS.444..290B} to obtain the scale height and zero point of a
Galactic latitude distribution. For the PNe we find a scale height of
0\fdeg.92\,$\pm$\,0\fdeg.11 with a zero point at
$b$\,$=$\,$-$0\fdeg.01\,$\pm$\,0\fdeg.01. However, due to our limited latitude
coverage, this scale height should be taken as a lower limit. Indeed there is a
38.9\,\% KS-test probability that the distribution of PNe across the GP in our
survey area is drawn from a homogeneously distributed sample. The vertical zero
point of the PN distribution coincides (within the uncertainties) with the GP.
This shows that the PNe trace an older, evolved population of objects. For the
Jets and PDRs the scale heights are smaller with 0\fdeg.65\,$\pm$\,0\fdeg.06 and
0\fdeg.66\,$\pm$\,0\fdeg.04, respectively. The distribution zero points are at
$-$0\fdeg.18\,$\pm$\,0\fdeg.01 and $-$0\fdeg.17\,$\pm$\,0\fdeg.01, respectively.
Thus, within the uncertainties, the scale height and zero points of these
distributions are identical, even if a KS-test gives only a 34.1\,\% probability
that the vertical distributions of both groups are drawn from the same parent
distribution. They hence trace the same component of the star formation process.
The vertical zero points for groups of jets and unknown objects are
significantly below the GP. This is in good agreement with them tracing active
star formation, which coincides with the dust and young cluster distribution
which is shifted below the GP in the longitude range of our survey (e.g.
\citet{Drimmel2003}, \citet{Marshall2006}, \citet{2014MNRAS.444..290B}).

\subsection{Jets and Outflows, PDRs, Star Formation}

We can use all the \htwo\ features which are classified as Jets or PDRs as
indicators of star formation activity. Jets most commonly trace young, accreting
protostars and/or Classical T-Tauri stars. Objects we have classified as PDRs
are most likely excited by slightly more evolved young stars and are often found
near clusters or intermediate mass stars. 

Considering the respective survey areas, the Cygnus region clearly has the
highest projected Jet/PDR group density. While along the GP there are on average
about 2.15 Jet groups per square degree, in Cygnus the density is 5.00 per square
degree. For PDRs there are 4.43 groups per square degree along the GP and 8.41
in Cygnus. This is most likely due to the fact that we specifically targeted
high column density regions in Cygnus, i.e. active places of star formation.
However, we used the same strategy in Auriga, which can be considered an example
SF region in the outer GP, but we find on average only 1.44 Jet groups and 0.87
PDRs per square degree. This is a clear indication that the number of \htwo\
features is related to the general star formation activity which seems lowest in
Auriga despite the bias in the survey. Furthermore, both SF indicators (Jets and
PDRs) are clearly more prevalent in the inner GP compared to the outer GP. 

A similar picture emerges if one uses the total flux of all features in each
part of the survey to trace star formation activity. All the details are
summarised in Table\,\ref{regionresulttable}. The total \htwo\ flux per square
degree from Jets is about 6.1 times higher in Cygnus compared to the GP. The
flux per area for PDRs is 3.0 times higher in Cygnus than in the GP. In Auriga
the flux per square degree from jets is 6.6 times lower than in the GP and for
PDRs its 4.5 times lower. Thus, despite the focus on SF regions in the
observations of Auriga, this region clearly stands out as the least active SF
region in the survey.

We further investigate the median total fluxes for object groups, which we
calculate by summing up the total fluxes of all \htwo\ features in the group and
calculating the median over all groups (see Table\,\ref{regionaverages}). These
fluxes can be considered as typical fluxes for each group of objects, as they
are not influenced by extreme outliers (see e.g. the two extremely bright SNRs
in Section\,\ref{snr_results}). For Jet groups and PDRs these fluxes are
generally lower in the outer GP than in the inner GP. For Jets the variations of
the median total fluxes in the four sub-regions are less than a factor of a few,
suggesting that the typical group of jets and outflows are similar in all the
investigated regions and that extinction and distances to the typical Jet groups
are comparable. Furthermore, the median total fluxes of the PDRs are very
similar in the inner GP and Cygnus/Auriga, i.e. the typical SF regions in these
areas are comparable in terms of their flux, but their numbers are highly
variable.

In summary, the Jet and PDR features in the survey give a clear indication of
the differences in the currently ongoing star formation activity (traced by
\htwo) in the areas covered by the survey. There are clearly more Jet groups and
PDRs per unit area in the more active star forming regions but the individual
objects are typically of a similar brightness. Please note that we cannot
discuss any  influences on the observed fluxes by systematic differences in
distance and extinction to the typical objects observed in the various parts of
the survey.

We are preparing more detailed investigations of the jets and outflows in Cygnus
and Auriga. Several regions along the GP have also already been studied in
detail (e.g. \citet{2012MNRAS.421.3257I}, \citet{2012MNRAS.425.1380I},
\citet{2011MNRAS.418.1375F}, \citet{2012ApJS..200....2L},
\citet{2012ApJ...756..151D}, \citet{2012AJ....144..151L},
\citet{2013MNRAS.429.1386D}, \citet{2013ApJS..208...23L},
\citet{2015MNRAS.446.2640D}, \citet{2015ApJ...803..100D}).

\subsection{Planetary Nebulae}
 
A list of groups of \htwo\ detections which coincide with known PNe or which we
consider to be new PNe candidates is given in Table\,\ref{pndatatable} in the
Appendix. We list the group/source ID (which contains the Galactic coordinates),
Right Ascension and Declination, the radius of the circumscribing circle
enclosing all detected \htwo\ emission, the area of emission, and the
corresponding total and median fluxes. Finally, for previously known objects
considered to be PNe, we give the PN\,G identifier where available. In cases
where a known object is listed in SIMBAD but has not been identified as a PN
then we give the alternative identifier (e.g. the IRAS name). 

Approximately 60\,\% of the \htwo\ detections in Table\,\ref{pndatatable}
correspond to emission features that have no corresponding source in SIMBAD and
have not been identified as PN or PN candidates in the literature. We list these
as {\it New} and have flagged them as possible PNe on the basis of their
morphology and lack of association with known star formation activity. We stress
that these are candidate PNe and their true nature will be established by
follow-up observations (Gledhill et al. in prep.). These include two PNe
candidates that have previously been identified as MHOs in
\citep{2012MNRAS.421.3257I}.

The number of PNe per square degree is 1.25 for the GP, with 1.18 and 1.31 per
square degree in the inner and outer GP regions respectively (defined as
$-3\degr < l < 30\degr$ and $30\degr \leq l < 66\degr$). Interestingly, the
higher space density in the outer GP arises from {\it New} detections; 58\,\% of
\htwo\ detections in the inner GP are new, and 65\,\% in the outer GP,
corresponding to 0.68 and 0.85 objects per square degree, respectively. By
contrast, the density of previously known PNe (with PN\,G identifiers) which
also have \htwo\ emission, is 0.34 per square degree in both regions.

The lower space density of PNe in the inner GP may be a consequence of increased
extinction along these sightlines. This is further supported by the larger
median flux for inner GP PNe ($570 \times 10^{-19}$\,W\,m$^{-2}$ compared with
$366 \times 10^{-19}$\,W\,m$^{-2}$) suggesting that we are sampling shorter
sightlines. However, the higher fraction of {\it New} \htwo-detected PNe in the
outer, compared to inner GP (65\,\% compared to 58\,\%) indicates that we have
not uncovered a population of inner-Galaxy PNe that were previously obscured in
optical surveys. The Galactic distribution of \htwo-detected PNe shown in
Fig.\,\ref{objdist} is actually similar to that of optically detected IPHAS PNe
\citep{2014MNRAS.443.3388S}.

\begin{table*}

\caption{\label{SNRlist} List of SNRs with identified extended \htwo\ emission
line features in our survey. The data (positions, size, type, 1\,GHz flux,
spectral index  and other identifiers) are taken from 
\citet{2014BASI...42...47G}. The area covered by \htwo\ emission, the total and
median fluxes as well as the number of individual \htwo\ emission regions are
also listed. Note that G\,6.5$-$0.4 also overlaps with several extended \htwo\
features. However, due to their visual appearance we attribute all of these to
W\,28.}

\renewcommand{\tabcolsep}{4pt}
\begin{center}
\begin{tabular}{cccccccccccp{1.2cm}}
\hline
Name & RA & DEC & Size & Area & 1 GHz flux & spectral & F$_{\rm tot}$ & F$_{\rm med}$ & number of & type & other \\
 & \multicolumn{2}{c}{(J2000)} & [arcmin] & [arcmin$^{2}$] & [Jy] & index & \multicolumn{2}{c}{[10$^{-15}$\,W\,m$^{-2}$]} & regions & & name \\
 \hline
  G1.0$-$0.1 & 17:48:30 & $-$28:09 & 8      & 0.12  & 15 & 0.6? & 0.24 & 0.22 & 4 &             S  & \\
  G1.4$-$0.1 & 17:49:39 & $-$27:46 & 10     & 0.076 & 2? & ? & 0.43 & 0.18 & 7 &             S  & \\
  G5.5$+$0.3 & 17:57:04 & $-$24:00 & 15x12  & 0.28  & 5.5 & 0.7 & 1.12 & 0.82 & 8 &           S  & \\
  G6.1$+$0.5 & 17:57:29 & $-$23:25 & 18x12  & 0.052 & 4.5 & 0.9 & 0.11 & 0.10 & 3 &              S  & \\
  G6.4$-$0.1 & 18:00:30 & $-$23:26 & 48     & 34.8  & 310 & varies & 126 & 89 & 1530 &  C  & W\,28 \\
  G9.9$-$0.8 & 18:10:41 & $-$20:43 & 12     & 0.11  & 6.7 & 0.4 & 0.10 & 0.09 & 20 &              S  & \\
 G11.2$-$0.3 & 18:11:27 & $-$19:25 &  4     & 1.70  & 22 & 0.5 & 5.2 & 3.2 & 77 &         C  & \\
 G13.5$+$0.2 & 18:14:14 & $-$17:12 & 5x4    & 0.049 & 3.5? & 1.0? & 0.06 & 0.05 & 6 &               S  & \\
 G16.0$-$0.5 & 18:21:56 & $-$15:14 & 15x10  & 0.40  & 2.7 & 0.6 & 1.6 & 0.54 & 47 &          S  & \\
 G18.1$-$0.1 & 18:24:34 & $-$13:11 & 8      & 0.34  & 4.6 & 0.5 & 3.0 & 0.92 & 48 &          S  & \\
 G18.9$-$1.1 & 18:29:50 & $-$12:58 & 33     & 0.80  & 37 & 0.39 & 1.1 & 0.97 & 102 &         C? & \\
 G21.6$-$0.8 & 18:33:40 & $-$10:25 & 13     & 0.020 & 1.4 & 0.5? & 0.02 & 0.02 & 5 &                S  & \\
 G21.8$-$0.6 & 18:32:45 & $-$10:08 & 20     & 1.75  & 65 & 0.56 & 4.3 & 3.7 & 119 &        S  & Kes\,69 \\
 G24.7$+$0.6 & 18:34:10 & $-$07:05 & 30x15  & 0.42  & 20? & 0.2? & 0.48 & 0.44 & 70 &           C? & \\
 G27.4$+$0.0 & 18:41:19 & $-$04:56 & 4      & 0.054 & 6 & 0.68 & 0.09 & 0.09 & 9 & S  &               4C$-$04.71 \\
 G27.8$+$0.6 & 18:39:50 & $-$04:24 & 50x30  & 0.11  & 30 & varies & 0.15 & 0.14 & 3 &               F & \\
 G28.8$+$1.5 & 18:39:00 & $-$02:55 & 100?   & 0.046 & ? & 0.4? & 0.04 & 0.04 & 7 &               S? & \\
 G31.9$+$0.0 & 18:49:25 & $-$00:55 & 7x5    & 2.32  & 25 & varies & 15.6 & 7.2 & 102 &       S  & 3C391 \\
 G32.1$-$0.9 & 18:53:10 & $-$01:08 & 40?    & 0.55  & ? & ? & 1.8 & 0.70 & 71 &          C? & \\
 G32.8$-$0.1 & 18:51:25 & $-$00:08 & 17     & 2.20  & 11? & 0.2? & 7.5 & 2.9 & 203 &        S? & Kes\,78 \\
 G33.2$-$0.6 & 18:53:50 & $-$00:02 & 18     & 0.12  & 3.5 & varies & 0.71 & 0.17 & 12 &            S  & \\
 G34.7$-$0.4 & 18:56:00 & $+$01:22 & 35x27  & 62.9  & 250 & 0.37 & 263 & 157 & 2852 & C  & W\,44 \\
 G38.7$-$1.3 & 19:06:40 & $+$04:28 & 32x19? & 0.26  & ? & ? & 0.57 & 0.29 & 43 &            S  & \\
 G39.2$-$0.3 & 19:04:08 & $+$05:28 & 8x6    & 0.52  & 18 & 0.34 & 0.75 & 0.70 & 49 &           C  & 3C396 \\
 G43.3$-$0.2 & 19:11:08 & $+$09:06 & 4x3    & 5.0   & 38 & 0.46 & 15.5 & 13.7 & 107 &     S  & W\,49B \\
 G54.4$-$0.3 & 19:33:20 & $+$18:56 & 40     & 0.39  & 28 & 0.5 & 0.52 & 0.41 & 54 &           S  & HC\,40 \\
 G65.1$+$0.6 & 19:54:40 & $+$28:35 & 90x50  & 0.13  & 5.5 & 0.61 & 0.06 & 0.09 & 7 &               S  & \\
G357.7$+$0.3 & 17:38:35 & $-$30:44 & 24     & 0.032 & 10 & 0.4? & 0.09 & 0.08 & 2 &               S  & \\
G359.0$-$0.9 & 17:46:50 & $-$30:16 & 23     & 0.15  & 23 & 0.5 & 0.30 & 0.30 & 10 &            S  & \\
G359.1$-$0.5 & 17:45:30 & $-$29:57 & 24     & 1.31  & 14 & 0.4? & 10.3 & 3.3 & 53 &         S  & \\

\hline
\end{tabular}
\end{center}
\end{table*}

\subsection{Supernova Remnants}\label{snr_results}
 
There are about 300 known SNRs in the Milky Way \citep{2014BASI...42...47G}, and
119 SNRs are either fully or partially covered in the UWISH2 survey, including
seven SNRs in Cygnus and one SNR in Auriga. We have detected \htwo\ emission
features which are most likely associated with SNRs for 30 of them. Hence, the
\htwo\ SNR detection rate is 25\,\%. Table\,\ref{SNRlist} lists the SNRs with
\htwo\ emission features where we list the SNR name, coordinates, sizes, types
and other names. All parameters in this Table (except area covered in \htwo\ and
fluxes) are from \citet{2014BASI...42...47G}. Please note that the SNR
G6.5$-$0.4 also overlaps with several extended \htwo\ features. However, due to
their visual appearance we attribute all of these to the larger, more extended
SNR W\,28.

The SNRs bright in \htwo\ emission in Table\,\ref{SNRlist}, e.g., W\,28, 3C391,
W\,44, and W\,49B, are prototypical SNRs interacting with molecular clouds. In
these SNRs, the \htwo\ features fall on bright radio filaments, where the SN
blast wave might be encountering dense environment. These \htwo\ emission
features are probably shock excited. In some SNRs, however, \htwo\ features are
just outside the radio continuum boundary (e.g. in G\,11.2$-$0.3), and those
features could be radiatively excited \citep{2014IAUS..296..214K}. Note that
W\,44 and W\,28 are responsible for 84\,\% of the total \htwo\ emission
associated with SNRs in our survey (57\,\% and 27\,\%, respectively). A much
more detailed discussion of the \htwo\ emission features associated with SNRs
will be presented in a forthcoming paper (Lee et al. 2015, in preparation). 

\section{Conclusions}

We have used WFCAM at UKIRT to conduct a large survey for emission of the \htwo\
1\,--\,0\,S(1) line at 2.122\,$\mu$m. An unbiased survey along the GP from $l
\approx 357\degr$ to $l \approx 65\degr$ and $|b| \le $\,1\fdeg.5 covers about
209 square degrees. We have further targeted high column density areas in Cygnus
(42.0 square degrees) and Auriga (35.5 square degrees). 

We have compiled a catalogue of extended \htwo\ emission line features in this
survey. All features were automatically detected and manually verified. We
estimate that only 1\,--\,2\,\% of the objects in the catalogue are false
positives and that 95\,\% of the real automatic detections are in the final
catalogue. Mostly small features in the vicinity of larger extended \htwo\
emission might be missing but these do not contribute with any significance to
the total detected \htwo\ flux. All features are also manually classified as
either part of a Jet/outflow, PN or SNR. All other objects are unclassified but
these are most likely part of PDRs.

In total, 33200 individual extended \htwo\ emission line features are contained
in our catalogue. There are about 700 groups of jet/outflow features, 284 PNe,
30 SNRs and about 1300 groups of PDRs. The total \htwo\ flux is dominated by
the PDR and SNR features (each accounting for 40\,--\,45\,\% of all the flux).
The Jet groups and PNe each contain about 7\,--\,9\,\% of the flux.

We find that star formation (traced by \htwo\ emission of Jets and PDRs) is
strongest in the inner GP (less than 30\degr\ from the Galactic Centre) and in
the Cygnus region. The latter containing, due to our targeted survey, the
highest number of Jet or PDR groups per square degree. Auriga clearly shows the
lowest star formation activity based on all our measures (density of sources,
total \htwo\ flux etc.) and there is also a clear decline in the star formation
activity with distance from the Galactic Centre.

About 60\,\% of all the PNe and candidate PNe in our catalogue have no known
counterpart in any of the PNe catalogues. Hence our survey has uncovered a
significant, unknown population of young and or embedded PNe candidates in the
GP. Their spatial distribution, however, is very similar to the optically
detected PNe.

Of all SNRs (partially) covered by our survey, one quarter has detectable
\htwo\  emission. The total flux in these \htwo\ features is strongly dominated
by W\,44 and W\,28 which together contain 84\,\% of all the \htwo\ flux
associated with SNRs.

\section*{acknowledgements}
 
The UWISH2 and UWISH2-E survey team would like to acknowledge the UKIRT support
staff, particularly the Telescope System Specialists (Thor Wold, Tim Carroll and
Jack Ehle) and the many UKIRT observers who have obtained data for the project
via flexible scheduling. We would like to thank the Referee Paul Goldsmith for
his helpfull comments. We also acknowledge the Cambridge Astronomical Survey
Unit and the WFCAM Science Archive for the reduction and ingest of the survey
data.  B.-C.\,K. was supported by the National Research Foundation of Korea
(NRF) grant funded by the Korea Government (MSIP) (No.\,2012R1A4A1028713). The
United Kingdom Infrared Telescope is operated by the Joint Astronomy Centre on
behalf of the Science and Technology Facilities Council of the U.K. Finally, we
also thank the UKIRT Time Allocation Committee for their support of this
long-term project. This research made use of Montage, funded by the National
Aeronautics and Space Administration's Earth Science Technology Office,
Computation Technologies Project, under Cooperative Agreement Number NCC5-626
between NASA and the California Institute of Technology. Montage is maintained
by the NASA/IPAC Infrared Science Archive.

\bibliographystyle{mn2e}
\bibliography{references}

\begin{thebibliography}{}

\bibitem[\protect\citeauthoryear{{Bally}, {Ginsburg}, {Probst}, {Reipurth},
  {Shirley} \& {Stringfellow}}{{Bally} et~al.}{2014}]{2014AJ....148..120B}
{Bally} J.,  {Ginsburg} A.,  {Probst} R.,  {Reipurth} B.,  {Shirley} Y.~L.,
  {Stringfellow} G.~S.,  2014, \aj, 148, 120

\bibitem[\protect\citeauthoryear{{Benjamin}, {Churchwell}, {Babler}, {Bania},
  {Clemens}, {Cohen}, {Dickey} \& {et al.}}{{Benjamin}
  et~al.}{2003}]{2003PASP..115..953B}
{Benjamin} R.~A.,  {Churchwell} E.,  {Babler} B.~L.,  {Bania} T.~M.,  {Clemens}
  D.~P.,  {Cohen} M.,  {Dickey} J.~M.,    {et al.} 2003, \pasp, 115, 953

\bibitem[\protect\citeauthoryear{{Buckner} \& {Froebrich}}{{Buckner} \&
  {Froebrich}}{2014}]{2014MNRAS.444..290B}
{Buckner} A.~S.~M.,  {Froebrich} D.,  2014, \mnras, 444, 290

\bibitem[\protect\citeauthoryear{{Churchwell}, {Babler}, {Meade}, {Whitney},
  {Benjamin}, {Indebetouw}, {Cyganowski}, {Robitaille}, {Povich}, {Watson} \&
  {Bracker}}{{Churchwell} et~al.}{2009}]{2009PASP..121..213C}
{Churchwell} E.,  {Babler} B.~L.,  {Meade} M.~R.,  {Whitney} B.~A.,  {Benjamin}
  R.,  {Indebetouw} R.,  {Cyganowski} C.,  {Robitaille} T.~P.,  {Povich} M.,
  {Watson} C.,    {Bracker} S.,  2009, \pasp, 121, 213

\bibitem[\protect\citeauthoryear{{Cohen}, {Wheaton} \& {Megeath}}{{Cohen}
  et~al.}{2003}]{2003AJ....126.1090C}
{Cohen} M.,  {Wheaton} W.~A.,    {Megeath} S.~T.,  2003, \aj, 126, 1090

\bibitem[\protect\citeauthoryear{{Davis} \& {Eisloeffel}}{{Davis} \&
  {Eisloeffel}}{1995}]{1995A&A...300..851D}
{Davis} C.~J.,  {Eisloeffel} J.,  1995, \aap, 300, 851

\bibitem[\protect\citeauthoryear{{Davis}, {Froebrich}, {Stanke}, {Megeath},
  {Kumar}, {Adamson}, {Eisl{\"o}ffel}, {Gredel}, {Khanzadyan}, {Lucas}, {Smith}
  \& {Varricatt}}{{Davis} et~al.}{2009}]{2009A&A...496..153D}
{Davis} C.~J.,  {Froebrich} D.,  {Stanke} T.,  {Megeath} S.~T.,  {Kumar}
  M.~S.~N.,  {Adamson} A.,  {Eisl{\"o}ffel} J.,  {Gredel} R.,  {Khanzadyan} T.,
   {Lucas} P.,  {Smith} M.~D.,    {Varricatt} W.~P.,  2009, \aap, 496, 153

\bibitem[\protect\citeauthoryear{{Dewangan}, {Mayya}, {Luna} \&
  {Ojha}}{{Dewangan} et~al.}{2015}]{2015ApJ...803..100D}
{Dewangan} L.~K.,  {Mayya} Y.~D.,  {Luna} A.,    {Ojha} D.~K.,  2015, \apj,
  803, 100

\bibitem[\protect\citeauthoryear{{Dewangan} \& {Ojha}}{{Dewangan} \&
  {Ojha}}{2013}]{2013MNRAS.429.1386D}
{Dewangan} L.~K.,  {Ojha} D.~K.,  2013, \mnras, 429, 1386

\bibitem[\protect\citeauthoryear{{Dewangan}, {Ojha}, {Anandarao}, {Ghosh} \&
  {Chakraborti}}{{Dewangan} et~al.}{2012}]{2012ApJ...756..151D}
{Dewangan} L.~K.,  {Ojha} D.~K.,  {Anandarao} B.~G.,  {Ghosh} S.~K.,
  {Chakraborti} S.,  2012, \apj, 756, 151

\bibitem[\protect\citeauthoryear{{Dewangan}, {Ojha}, {Grave} \&
  {Mallick}}{{Dewangan} et~al.}{2015}]{2015MNRAS.446.2640D}
{Dewangan} L.~K.,  {Ojha} D.~K.,  {Grave} J.~M.~C.,    {Mallick} K.~K.,  2015,
  \mnras, 446, 2640

\bibitem[\protect\citeauthoryear{{Drimmel}, {Cabrera-Lavers} \&
  {L{\'o}pez-Corredoira}}{{Drimmel} et~al.}{2003}]{Drimmel2003}
{Drimmel} R.,  {Cabrera-Lavers} A.,    {L{\'o}pez-Corredoira} M.,  2003, \aap,
  409, 205

\bibitem[\protect\citeauthoryear{{Dye}, {Warren}, {Hambly}, {Cross}, {Hodgkin},
  {Irwin}, {Lawrence} \& {et al.}}{{Dye} et~al.}{2006}]{2006MNRAS.372.1227D}
{Dye} S.,  {Warren} S.~J.,  {Hambly} N.~C.,  {Cross} N.~J.~G.,  {Hodgkin}
  S.~T.,  {Irwin} M.~J.,  {Lawrence} A.,    {et al.} 2006, \mnras, 372, 1227

\bibitem[\protect\citeauthoryear{{Froebrich}, {Davis}, {Ioannidis}, {Gledhill},
  {Takami}, {Chrysostomou}, {Drew} \& {et al.}}{{Froebrich}
  et~al.}{2011}]{2011MNRAS.413..480F}
{Froebrich} D.,  {Davis} C.~J.,  {Ioannidis} G.,  {Gledhill} T.~M.,  {Takami}
  M.,  {Chrysostomou} A.,  {Drew}   {et al.} 2011, \mnras, 413, 480

\bibitem[\protect\citeauthoryear{{Froebrich} \& {Ioannidis}}{{Froebrich} \&
  {Ioannidis}}{2011}]{2011MNRAS.418.1375F}
{Froebrich} D.,  {Ioannidis} G.,  2011, \mnras, 418, 1375

\bibitem[\protect\citeauthoryear{{Giannini}, {McCoey}, {Caratti o Garatti},
  {Nisini}, {Lorenzetti} \& {Flower}}{{Giannini}
  et~al.}{2004}]{2004A&A...419..999G}
{Giannini} T.,  {McCoey} C.,  {Caratti o Garatti} A.,  {Nisini} B.,
  {Lorenzetti} D.,    {Flower} D.~R.,  2004, \aap, 419, 999

\bibitem[\protect\citeauthoryear{{Giannini}, {Nisini}, {Caratti o Garatti} \&
  {Lorenzetti}}{{Giannini} et~al.}{2002}]{2002ApJ...570L..33G}
{Giannini} T.,  {Nisini} B.,  {Caratti o Garatti} A.,    {Lorenzetti} D.,
  2002, \apjl, 570, L33

\bibitem[\protect\citeauthoryear{{Green}}{{Green}}{2009}]{2009BASI...37...45G}
{Green} D.~A.,  2009, Bulletin of the Astronomical Society of India, 37, 45

\bibitem[\protect\citeauthoryear{{Green}}{{Green}}{2014}]{2014BASI...42...47G}
{Green} D.~A.,  2014, Bulletin of the Astronomical Society of India, 42, 47

\bibitem[\protect\citeauthoryear{{Hartigan}, {Reiter}, {Smith} \&
  {Bally}}{{Hartigan} et~al.}{2015}]{2015AJ....149..101H}
{Hartigan} P.,  {Reiter} M.,  {Smith} N.,    {Bally} J.,  2015, \aj, 149, 101

\bibitem[\protect\citeauthoryear{{Hewett}, {Warren}, {Leggett} \&
  {Hodgkin}}{{Hewett} et~al.}{2006}]{2006MNRAS.367..454H}
{Hewett} P.~C.,  {Warren} S.~J.,  {Leggett} S.~K.,    {Hodgkin} S.~T.,  2006,
  \mnras, 367, 454

\bibitem[\protect\citeauthoryear{{Ioannidis} \& {Froebrich}}{{Ioannidis} \&
  {Froebrich}}{2012a}]{2012MNRAS.421.3257I}
{Ioannidis} G.,  {Froebrich} D.,  2012a, \mnras, 421, 3257

\bibitem[\protect\citeauthoryear{{Ioannidis} \& {Froebrich}}{{Ioannidis} \&
  {Froebrich}}{2012b}]{2012MNRAS.425.1380I}
{Ioannidis} G.,  {Froebrich} D.,  2012b, \mnras, 425, 1380

\bibitem[\protect\citeauthoryear{{Koo}}{{Koo}}{2014}]{2014IAUS..296..214K}
{Koo} B.-C.,  2014, in {Ray} A.,  {McCray} R.~A.,  eds, IAU Symposium Vol.~296
  of IAU Symposium, {Infrared [Fe II] and Dust Emissions from Supernova
  Remnants}.
pp 214--221

\bibitem[\protect\citeauthoryear{{Lawrence}, {Warren}, {Almaini}, {Edge},
  {Hambly}, {Jameson}, {Lucas} \& {et al.}}{{Lawrence}
  et~al.}{2007}]{2007MNRAS.379.1599L}
{Lawrence} A.,  {Warren} S.~J.,  {Almaini} O.,  {Edge} A.~C.,  {Hambly} N.~C.,
  {Jameson} R.~F.,  {Lucas} P.,    {et al.} 2007, \mnras, 379, 1599

\bibitem[\protect\citeauthoryear{{Lee}, {Liao}, {Froebrich}, {Karr},
  {Ioannidis}, {Lee}, {Su}, {Liu}, {Duan} \& {Takami}}{{Lee}
  et~al.}{2013}]{2013ApJS..208...23L}
{Lee} H.-T.,  {Liao} W.-T.,  {Froebrich} D.,  {Karr} J.,  {Ioannidis} G.,
  {Lee} Y.-H.,  {Su} Y.-N.,  {Liu} S.-Y.,  {Duan} H.-Y.,    {Takami} M.,  2013,
  \apjs, 208, 23

\bibitem[\protect\citeauthoryear{{Lee}, {Takami}, {Duan}, {Karr}, {Su}, {Liu},
  {Froebrich} \& {Yeh}}{{Lee} et~al.}{2012}]{2012ApJS..200....2L}
{Lee} H.-T.,  {Takami} M.,  {Duan} H.-Y.,  {Karr} J.,  {Su} Y.-N.,  {Liu}
  S.-Y.,  {Froebrich} D.,    {Yeh} C.~C.,  2012, \apjs, 200, 2

\bibitem[\protect\citeauthoryear{{Lee}, {Koo}, {Lee}, {Lee}, {Shinn}, {Kim},
  {Kim}, {Pyo}, {Moon}, {Yoon}, {Chun}, {Froebrich}, {Davis}, {Varricatt},
  {Kyeong}, {Hwang}, {Park}, {Lee}, {Lee} \& {Ishiguro}}{{Lee}
  et~al.}{2014}]{2014MNRAS.443.2650L}
{Lee} J.-J.,  {Koo} B.-C.,  {Lee} Y.-H.,  {Lee} H.-G.,  {Shinn} J.-H.,  {Kim}
  H.-J.,  {Kim} Y.,  {Pyo} T.-S.,  {Moon} D.-S.,  {Yoon} S.-C.,  {Chun} M.-Y.,
  {Froebrich} D.,  {Davis} C.~J.,  {Varricatt} W.~P.,  {Kyeong} J.,  {Hwang}
  N.,  {Park} B.-G.,  {Lee} M.~G.,  {Lee} H.~M.,    {Ishiguro} M.,  2014,
  \mnras, 443, 2650

\bibitem[\protect\citeauthoryear{{Lim}, {Lyo}, {Kim} \& {Byun}}{{Lim}
  et~al.}{2012}]{2012AJ....144..151L}
{Lim} W.,  {Lyo} A.-R.,  {Kim} K.-T.,    {Byun} D.-Y.,  2012, \aj, 144, 151

\bibitem[\protect\citeauthoryear{{Lucas}, {Hoare}, {Longmore}, {Schr{\"o}der},
  {Davis}, {Adamson}, {Bandyopadhyay} \& {et al.}}{{Lucas}
  et~al.}{2008}]{Lucas2008}
{Lucas} P.~W.,  {Hoare} M.~G.,  {Longmore} A.,  {Schr{\"o}der} A.~C.,  {Davis}
  C.~J.,  {Adamson} A.,  {Bandyopadhyay} R.~M.,    {et al.} 2008, \mnras, 391,
  136

\bibitem[\protect\citeauthoryear{{Marshall}, {Robin}, {Reyl{\'e}}, {Schultheis}
  \& {Picaud}}{{Marshall} et~al.}{2006}]{Marshall2006}
{Marshall} D.~J.,  {Robin} A.~C.,  {Reyl{\'e}} C.,  {Schultheis} M.,
  {Picaud} S.,  2006, \aap, 453, 635

\bibitem[\protect\citeauthoryear{{Miszalski}, {Parker}, {Acker}, {Birkby},
  {Frew} \& {Kovacevic}}{{Miszalski} et~al.}{2008}]{2008MNRAS.384..525M}
{Miszalski} B.,  {Parker} Q.~A.,  {Acker} A.,  {Birkby} J.~L.,  {Frew} D.~J.,
   {Kovacevic} A.,  2008, \mnras, 384, 525

\bibitem[\protect\citeauthoryear{{Molinari}, {Swinyard}, {Bally}, {Barlow},
  {Bernard}, {Martin}, {Moore} \& {et al.}}{{Molinari}
  et~al.}{2010}]{2010PASP..122..314M}
{Molinari} S.,  {Swinyard} B.,  {Bally} J.,  {Barlow} M.,  {Bernard} J.-P.,
  {Martin} P.,  {Moore} T.,    {et al.} 2010, \pasp, 122, 314

\bibitem[\protect\citeauthoryear{{Nisini}, {Caratti o Garatti}, {Giannini} \&
  {Lorenzetti}}{{Nisini} et~al.}{2002}]{2002A&A...393.1035N}
{Nisini} B.,  {Caratti o Garatti} A.,  {Giannini} T.,    {Lorenzetti} D.,
  2002, \aap, 393, 1035

\bibitem[\protect\citeauthoryear{{Parker}, {Acker}, {Frew}, {Hartley},
  {Peyaud}, {Ochsenbein}, {Phillipps}, {Russeil}, {Beaulieu}, {Cohen},
  {K{\"o}ppen}, {Miszalski}, {Morgan}, {Morris}, {Pierce} \&
  {Vaughan}}{{Parker} et~al.}{2006}]{2006MNRAS.373...79P}
{Parker} Q.~A.,  {Acker} A.,  {Frew} D.~J.,  {Hartley} M.,  {Peyaud} A.~E.~J.,
  {Ochsenbein} F.,  {Phillipps} S.,  {Russeil} D.,  {Beaulieu} S.~F.,  {Cohen}
  M.,  {K{\"o}ppen} J.,  {Miszalski} B.,  {Morgan} D.~H.,  {Morris} R.~A.~H.,
  {Pierce} M.~J.,    {Vaughan} A.~E.,  2006, \mnras, 373, 79

\bibitem[\protect\citeauthoryear{{Sabin}, {Parker}, {Corradi},
  {Guzman-Ramirez}, {Morris}, {Zijlstra}, {Boji{\v c}i{\'c}} \& {et
  al.}}{{Sabin} et~al.}{2014}]{2014MNRAS.443.3388S}
{Sabin} L.,  {Parker} Q.~A.,  {Corradi} R.~L.~M.,  {Guzman-Ramirez} L.,
  {Morris} R.~A.~H.,  {Zijlstra} A.~A.,  {Boji{\v c}i{\'c}} I.~S.,    {et al.}
  2014, \mnras, 443, 3388

\bibitem[\protect\citeauthoryear{{Schmeja}, {Kumar} \& {Ferreira}}{{Schmeja}
  et~al.}{2008}]{2008MNRAS.389.1209S}
{Schmeja} S.,  {Kumar} M.~S.~N.,    {Ferreira} B.,  2008, \mnras, 389, 1209

\bibitem[\protect\citeauthoryear{{Stanke}, {McCaughrean} \&
  {Zinnecker}}{{Stanke} et~al.}{2002}]{2002A&A...392..239S}
{Stanke} T.,  {McCaughrean} M.~J.,    {Zinnecker} H.,  2002, \aap, 392, 239

\bibitem[\protect\citeauthoryear{{Varricatt}, {Davis}, {Ramsay} \&
  {Todd}}{{Varricatt} et~al.}{2010}]{2010MNRAS.404..661V}
{Varricatt} W.~P.,  {Davis} C.~J.,  {Ramsay} S.,    {Todd} S.~P.,  2010,
  \mnras, 404, 661

\bibitem[\protect\citeauthoryear{{Zhang}, {Fang}, {Wang}, {Sun}, {Wang},
  {Jiang} \& {Anathipindika}}{{Zhang} et~al.}{2015}]{2015arXiv150608372Z}
{Zhang} M.,  {Fang} M.,  {Wang} H.,  {Sun} J.,  {Wang} M.,  {Jiang} Z.,
  {Anathipindika} S.,  2015, ArXiv e-prints

\end{thebibliography}
\label{lastpage}

\clearpage
\newpage

\begin{appendix}

\section{{\tt map$\_$zp} distributions}
\clearpage
\newpage

\begin{figure}
\includegraphics[width=8.6cm,angle=0]{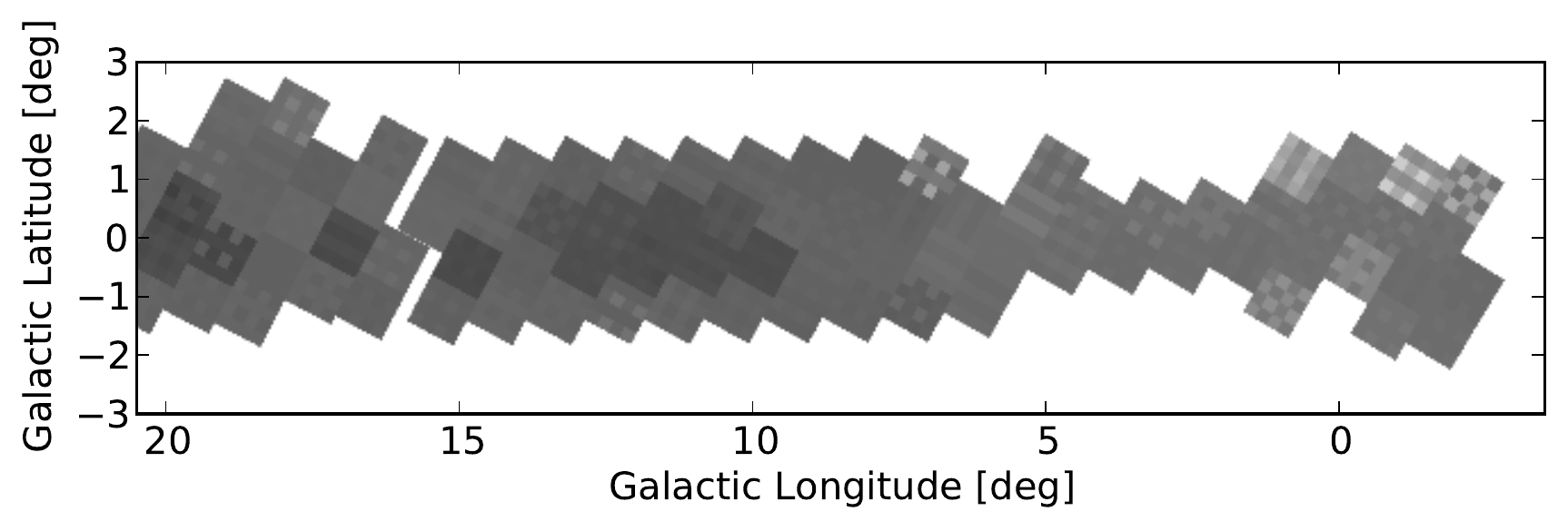} \\
\includegraphics[width=8.6cm,angle=0]{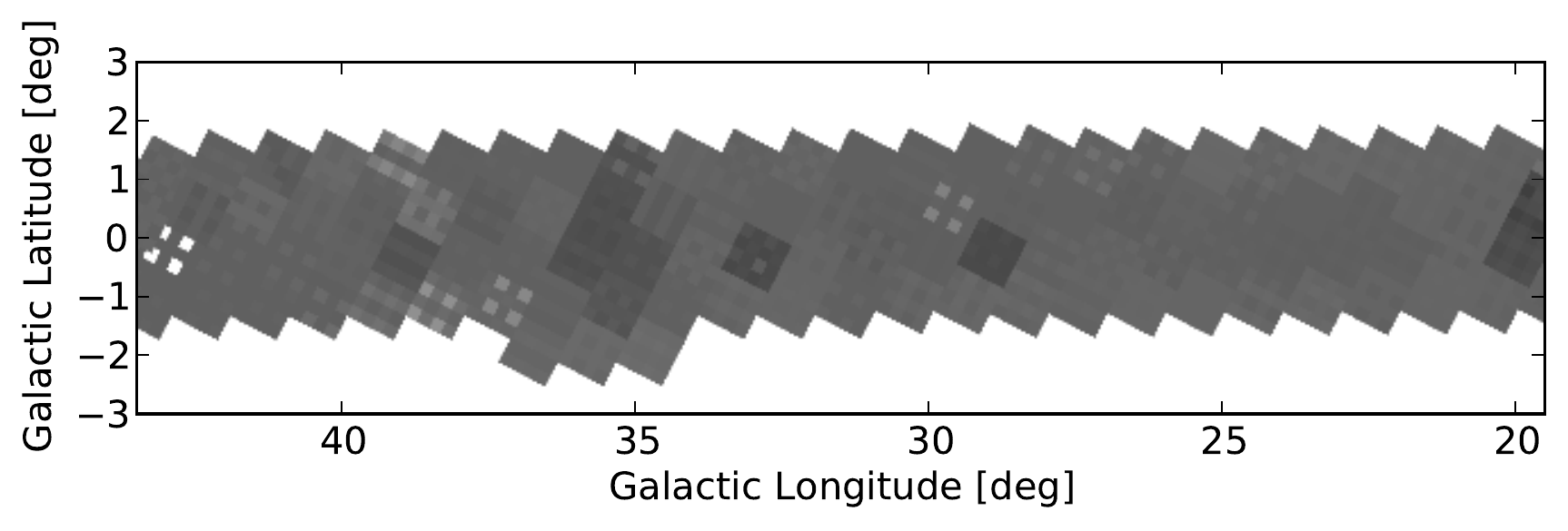} \\
\includegraphics[width=8.6cm,angle=0]{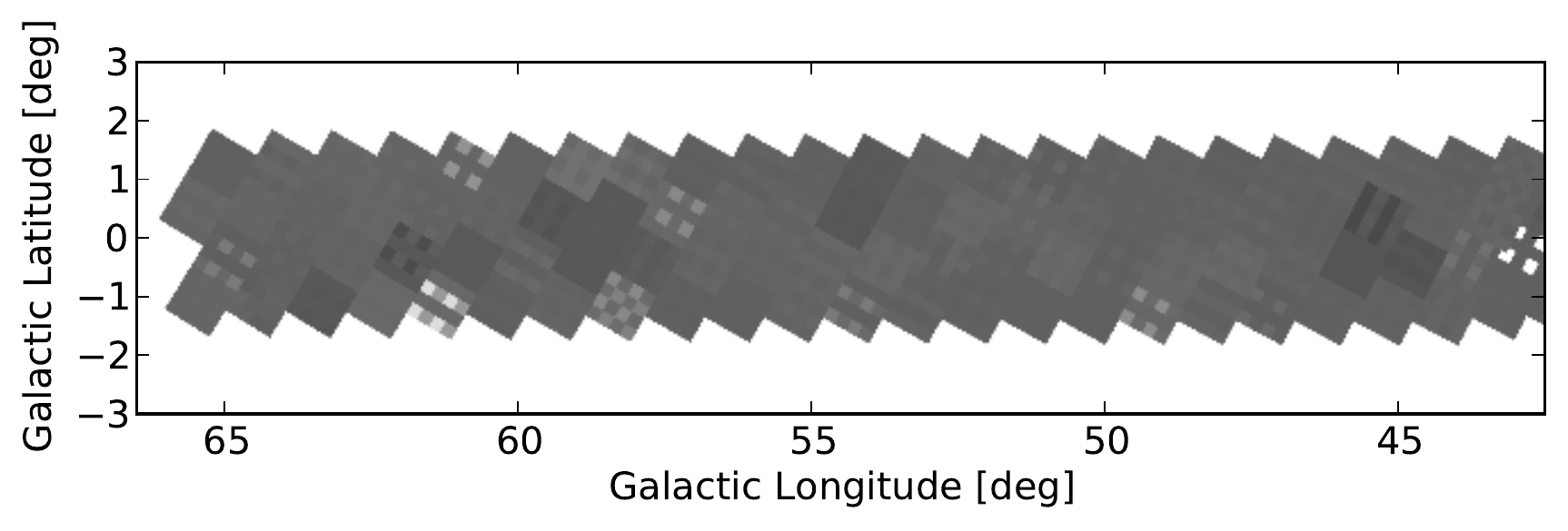} \\

\caption{\label{gpplotsapp}  Plots and {\tt mag$\_$zp} distribution in the
Galactic Plane area of the survey. Darker colours indicate higher values for
{\tt mag$\_$zp}.} 

\end{figure}

\begin{figure}
\includegraphics[width=8.6cm,angle=0]{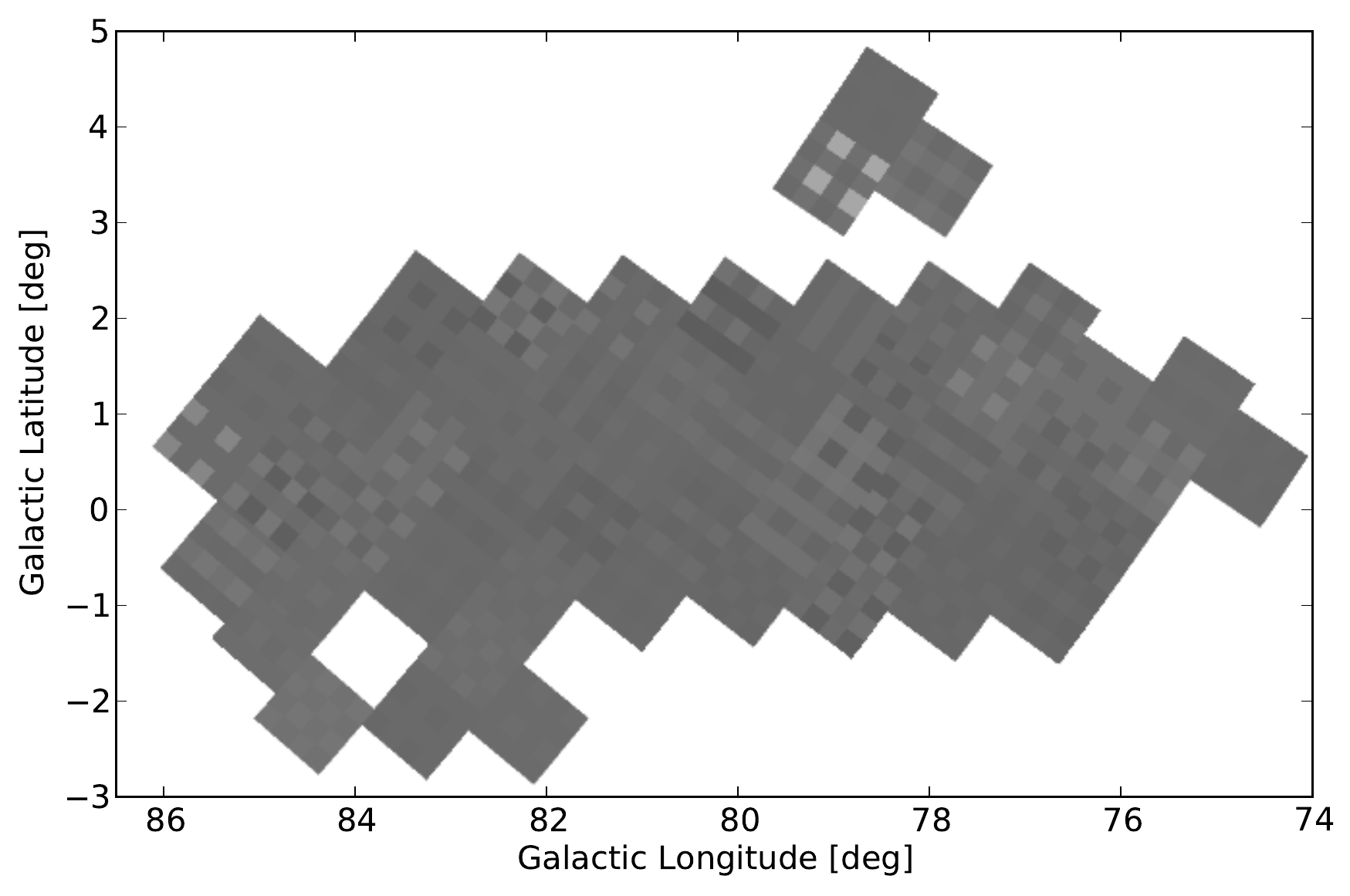}  \\
\includegraphics[width=8.6cm,angle=0]{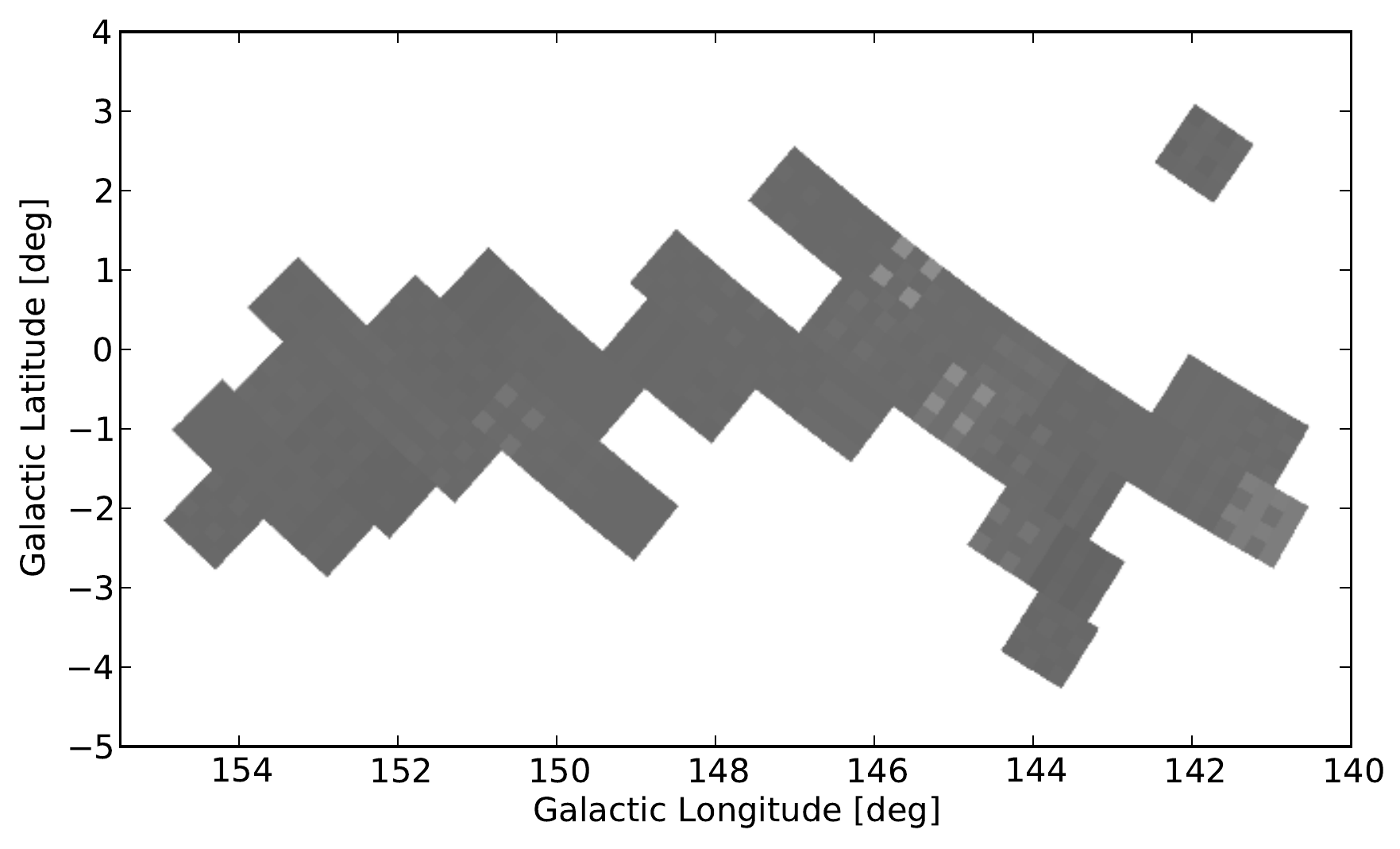} \\

\caption{\label{cygaurplotsapp}  As Fig.\,\ref{gpplotsapp} but for the Cygnus
(top) and Auriga (bottom) area of the survey.}

\end{figure}

\clearpage
\newpage

\section{Properties of PNe}
\clearpage
\newpage

\onecolumn

{\small
\renewcommand{\tabcolsep}{3pt}
\begin{center}
\begin{longtable}[l]{lccccccp{4.3cm}}

\caption{\label{pndatatable} Table listing all the PNe detected in our survey.
We list the UWISH2 PN source ID, which includes the Galactic coordinates, the RA
and DEC (J2000) positions of the geometric centre of the \htwo\ emitting
features in the PN, the radius around this central position enclosing all the
\htwo\ emission, the area covered with \htwo\ emission, the total and median
fluxes of the PN, and any other known name/identifier or coinciding object (not
necessarily a PN). 'New' in the last column indicates objects that have so far
not been recognised as PNe or PNe candidates in the literature.} \\

\hline
\multicolumn{1}{c}{Source ID} & RA & DEC & Radius & Area & F$_{tot}$ & F$_{med}$ & Other ID \\
 & [deg] & [deg] & [arcsec] & [arcsec$^{2}$] & \multicolumn{2}{c}{[10$^{-19}$Wm$^{-2}$]} &  \\ \hline
\endfirsthead

\multicolumn{8}{c}%
{{\bfseries \tablename\ \thetable{} -- continued from previous page}} \\
\hline 
\multicolumn{1}{c}{Source ID}  & RA & DEC & Radius & Area & F$_{tot}$ & F$_{med}$ & Other ID \\
 & [deg] & [deg] & [arcsec] & [arcsec$^{2}$] & \multicolumn{2}{c}{[10$^{-19}$Wm$^{-2}$]} &  \\ \hline
\endhead

\hline \multicolumn{8}{r}{{Continued on next page}} \\ \hline
\endfoot

\hline
\endlastfoot

PN$\_$UWISH2$\_$000.81878$-$0.04944 & 266.93718 & $-$28.26207 & 6.3 & 52.97 & 388.50 & 345.89 & SSTGC 841071 \\
PN$\_$UWISH2$\_$001.22588$+$0.56414 & 266.58223 & $-$27.59584 & 5.7 & 44.18 & 386.74 & 331.79 & New \\
PN$\_$UWISH2$\_$001.42213$-$0.61357 & 267.83773 & $-$28.03457 & 16.9 & 243.11 & 1706.78 & 1796.58 & New \\
PN$\_$UWISH2$\_$001.65056$+$0.18803 & 267.19290 & $-$27.42713 & 14.8 & 298.48 & 2544.15 & 2286.36 & PN G001.6$+$00.1 \\
PN$\_$UWISH2$\_$001.72196$-$0.82262 & 268.21435 & $-$27.88315 & 17.3 & 211.92 & 6355.37 & 1543.66 & New \\
PN$\_$UWISH2$\_$001.73118$+$0.44232 & 266.99457 & $-$27.22684 & 12.2 & 137.19 & 498.93 & 439.01 & New \\
PN$\_$UWISH2$\_$002.03824$-$0.34363 & 267.93092 & $-$27.36701 & 23.4 & 149.54 & 680.57 & 655.32 & New \\
PN$\_$UWISH2$\_$002.25319$+$0.55724 & 267.18642 & $-$26.72043 & 47.7 & 1312.63 & 1712.21 & 9032.45 & PN G002.2$+$00.5 \\
\end{longtable}
\end{center}
}
\twocolumn

\clearpage
\newpage

\onecolumn
\section{Images taken for the survey}
\clearpage
\newpage

\renewcommand{\tabcolsep}{3pt}
\begin{center}
\begin{longtable}{cccccccccc}

\caption{\label{imagedatatable} Table listing all the images taken for the
survey. We list the following columns: Tile name; Image name containing the date
of the observations; Right Ascension and Declination of image centre; Galactic
coordinates of the image centre; Seeing in the image; Magnitude zero point and
its uncertainty; Noise -- ($*$) in $10^{-19}$\,W\,m$^{-2}$\,arcsec$^{-2}$. This
Table is only an excerpt of the first few images. The full table will be made
available online.} \\

\hline
Tilename & Imagename & RA & DEC & l & b & Seeing & {\tt map$\_$zp} & $\Delta${\tt map$\_$zp} & Noise \\
  &   & \multicolumn{2}{c}{(J2000) [deg]} & \multicolumn{2}{c}{[deg]} & [arcsec] & [mag] & [mag] & ($*$) \\ \hline
\endfirsthead

\multicolumn{10}{c}%
{{\bfseries \tablename\ \thetable{} -- continued from previous page}} \\
\hline 
Tilename & Imagename & RA & DEC & l & b & Seeing & {\tt map$\_$zp} & $\Delta${\tt map$\_$zp} & Noise \\
  &   & \multicolumn{2}{c}{(J2000) [deg]} & \multicolumn{2}{c}{[deg]} & [arcsec] & [mag] & [mag] & ($*$) \\ \hline
\endhead

\hline \multicolumn{10}{r}{{Continued on next page}} \\ \hline
\endfoot

\hline
\endlastfoot

H2$\_$l15$\_$50$\_$284 & w20090728$\_$00340$\_$w & 284.66025 & $+$1.26845 & 34.88227 & $-$1.02369 &  0.66 & 21.22 &  0.020 &  2.78 \\
H2$\_$l15$\_$50$\_$284 & w20090728$\_$00340$\_$x & 285.10187 & $+$1.26744 & 35.08305 & $-$1.41693 &  0.69 & 21.22 &  0.020 &  2.97 \\
H2$\_$l15$\_$50$\_$284 & w20090728$\_$00340$\_$y & 285.10358 & $+$1.70699 & 35.47493 & $-$1.21764 &  0.72 & 21.22 &  0.020 &  2.87 \\
H2$\_$l15$\_$50$\_$284 & w20090728$\_$00340$\_$z & 284.66022 & $+$1.70904 & 35.27429 & $-$0.82248 &  0.68 & 21.22 &  0.020 &  2.99 \\
H2$\_$l15$\_$50$\_$284 & w20090728$\_$00363$\_$w & 284.66016 & $+$1.48862 & 35.07814 & $-$0.92308 &  0.63 & 21.23 &  0.051 &  2.68 \\
H2$\_$l15$\_$50$\_$284 & w20090728$\_$00363$\_$x & 285.10175 & $+$1.48762 & 35.27891 & $-$1.31624 &  0.67 & 21.23 &  0.051 &  2.81 \\
H2$\_$l15$\_$50$\_$284 & w20090728$\_$00363$\_$y & 285.10352 & $+$1.92713 & 35.67075 & $-$1.11699 &  0.69 & 21.23 &  0.051 &  2.78 \\
H2$\_$l15$\_$50$\_$284 & w20090728$\_$00363$\_$z & 284.66010 & $+$1.92916 & 35.47007 & $-$0.72185 &  0.65 & 21.23 &  0.051 &  2.97 \\
\end{longtable}
\end{center}

\clearpage
\newpage

\section{Complete \htwo\ Source List}
\clearpage
\newpage

\onecolumn

\begin{landscape}

{\tiny
\renewcommand{\tabcolsep}{4pt}
\begin{center}
\begin{longtable}[l]{cccccccccccccc}

\caption{\label{objectdatatable} Table listing all the extended \htwo\ features
in our survey. We list the following columns: Source ID which includes the
Galactic Coordinates; Right Ascension and Declination in (J2000) system; Area of
the \htwo\ feature; Radius of the \htwo\ feature; Median surface brightness;
Maximum surface brightness; one-pixel {\it rms} noise of the background; Total
flux of the \htwo\ feature; Relative flux uncertainty due to photometric
calibration; Classification (Jet, PN, SNR or unknown, most likely PDR); Tile
name object is on; Image name the object is on; Name of group the object is
associated with. This Table is only an excerpt of the first few objects. The
full table will be made available online.} \\

\hline
Source ID & RA & DEC & A & r & F$_{sb}^{med}$ & F$_{sb}^{max}$ & F$_\sigma$ & F$^{tot}$ & $\Delta F / F$ & $C$ & Tile & Image & Group \\
 & \multicolumn{2}{c}{(J2000) [deg]} & [arcsec$^2$] & [arcsec] & \multicolumn{3}{c}{[10$^{-19}$Wm$^{-2}$arcsec$^{-2}$]} & [10$^{-19}$Wm$^{-2}$] & [\%] &  & name & name & name \\ \hline
\endfirsthead

\multicolumn{14}{c}%
{{\bfseries \tablename\ \thetable{} -- continued from previous page}} \\
\hline 
Source ID & RA & DEC & A & r & F$_{sb}^{med}$ & F$_{sb}^{max}$ & F$_\sigma$ & F$^{tot}$ & $\Delta F / F$ & $C$ & Tile & Image & Group \\
 & \multicolumn{2}{c}{(J2000) [deg]} & [arcsec$^2$] & [arcsec] & \multicolumn{3}{c}{[10$^{-19}$Wm$^{-2}$arcsec$^{-2}$]} & [10$^{-19}$Wm$^{-2}$] & [\%] &  & name & name & name \\ \hline
\endhead

\hline \multicolumn{14}{r}{{Continued on next page}} \\ \hline
\endfoot

\hline
\endlastfoot

UWISH2$\_$000.00652$+$0.15598 & 266.25682 & $-$28.84930 & 121.51 & 18.4 & 4.26 & 103.44 & 6.13 & 520.11 & 7.9 & u & H2$\_$lm2$\_$15$\_$16 & w20130827$\_$00152$\_$x & PDR$\_$UWISH2$\_$359.99949$+$0.16099 \\ 
UWISH2$\_$000.01128$+$0.16305 & 266.25276 & $-$28.84155 & 105.35 & 12.3 & 4.17 & 236.49 & 6.21 & 447.16 & 7.9 & u & H2$\_$lm2$\_$15$\_$16 & w20130827$\_$00152$\_$x & PDR$\_$UWISH2$\_$359.99949$+$0.16099 \\ 
UWISH2$\_$000.01490$+$0.16272 & 266.25525 & $-$28.83864 & 22.97 & 5.3 & 2.39 & 18.70 & 6.31 & 55.68 & 7.9 & u & H2$\_$lm2$\_$15$\_$16 & w20130827$\_$00152$\_$x & PDR$\_$UWISH2$\_$359.99949$+$0.16099 \\ 
UWISH2$\_$000.01868$+$0.16127 & 266.25891 & $-$28.83617 & 107.44 & 10.3 & 5.54 & 5159.72 & 5.95 & 3916.77 & 7.9 & u & H2$\_$lm2$\_$15$\_$16 & w20130827$\_$00152$\_$x & PDR$\_$UWISH2$\_$359.99949$+$0.16099 \\ 
UWISH2$\_$000.02914$+$0.16027 & 266.26611 & $-$28.82777 & 45.03 & 5.5 & 3.32 & 227.03 & 6.14 & 122.05 & 7.9 & u & H2$\_$lm2$\_$15$\_$16 & w20130827$\_$00152$\_$x & PDR$\_$UWISH2$\_$359.99949$+$0.16099 \\ 
UWISH2$\_$000.03140$+$0.15965 & 266.26807 & $-$28.82617 & 7.45 & 3.9 & 2.85 & 52.26 & 5.87 & 20.02 & 7.9 & u & H2$\_$lm2$\_$15$\_$16 & w20130827$\_$00152$\_$x & PDR$\_$UWISH2$\_$359.99949$+$0.16099 \\ 
UWISH2$\_$000.03229$+$0.15892 & 266.26931 & $-$28.82579 & 5.45 & 2.1 & 4.54 & 15.35 & 4.81 & 24.07 & 7.9 & u & H2$\_$lm2$\_$15$\_$16 & w20130827$\_$00152$\_$x & PDR$\_$UWISH2$\_$359.99949$+$0.16099 \\ 
UWISH2$\_$000.03537$+$0.16025 & 266.26985 & $-$28.82247 & 34.13 & 5.8 & 2.80 & 43.78 & 5.62 & 109.51 & 7.9 & u & H2$\_$lm2$\_$15$\_$16 & w20130827$\_$00152$\_$x & PDR$\_$UWISH2$\_$359.99949$+$0.16099 \\
\end{longtable}
\end{center}
}
\end{landscape}
\twocolumn

\end{appendix}

\end{document}